# Universal emergence of spatially-modulated structures induced by flexo-antiferrodistortive coupling in multiferroics

Eugene A. Eliseev,[1] Sergei V. Kalinin,[2,*] Yijia Gu,[3] Maya D. Glinchuk,[1] Victoria Khist,[1] Albina Borisevich,[2] Venkatraman Gopalan,[3] Long-Qing Chen,[3] and Anna N. Morozovska,[4,†]

[1] Institute for Problems of Materials Science, National Academy of Sciences of Ukraine, 3, Krjijanovskogo, 03142 Kiev, Ukraine

[2] The Center for Nanophase Materials Sciences, Oak Ridge National Laboratory, Oak Ridge, TN 37831

[3] Department of Materials Science and Engineering, Pennsylvania State University, University Park, PA 16802, USA

[4] Institute of Physics, National Academy of Sciences of Ukraine, 41, pr. Nauki, 03028 Kiev, Ukraine

**Abstract**

We proved the existence of a universal flexo-antiferrodistortive coupling as a necessary complement to the well-known flexoelectric coupling. The coupling is universal for all antiferrodistortive systems and can lead to the formation of incommensurate, spatially-modulated phases in multiferroics. Our analysis can provide a self-consistent mesoscopic explanation for a broad range of modulated domain structures observed experimentally in multiferroics.

---

[*] sergei2@ornl.gov

[†] morozo@i.com.ua

# 1. Introduction

Multiferroics, materials with multiple coupled order parameters, have emerged as an important topic in condensed matter physics [1, 2] due to both their intriguing physical behaviors and a broad variety of novel physical applications they enable. The unique physical properties of multiferroics originated from the complex interactions among the structural, polar and magnetic long-range order parameters [3, 4]. For instance, biquadratic and linear magnetoelectric couplings lead to such intriguing effects as giant magnetoelectric tunability of multiferroics [5, 6]. Biquadratic coupling of the structural and polar order parameters, introduced by Haun [7], Salje et al [8], Balashova and Tagantsev [9], and Tagantsev et al [10], are responsible for the unusual behavior of the dielectric and polar properties in ferroelastics − quantum paraelectrics. Daraktchiev et al. [11] considered the influence of biquadratic coupling between polarization and magnetization on the structure of ferroelectric domain walls in multiferroics. Dieguez et al [12] found that the behavior of the structural order parameter at the domain walls of multiferroic $BiFeO_3$ determines their structure and energy. In this regard, new intriguing phenomena emerging in nanoscale phase-separated ferroics can be represented as extremely dense domain structures.

Nanoscale phase separation in materials ranging from giant magnetoresistive manganites [13, 14, 15], ferroelectric relaxors [16, 17] and morphotropic materials [18, 19, 20], martensites [21, 22] and birelaxors [23] remains one of the active topics of research in condensed matter physics. Experiment [24] revealed the existence of the incommensurate modulation at the structural domain boundaries in multiferroic $Bi_ySm_{1-y}FeO_3$. Antiferrodistortive [25] and superstructural dynamic antiferroelectric-antiferrodistortive [26] modulation have also recently been observed in multiferroic $EuTiO_3$.

There are also a wide variety of modulated domain structures observed experimentally at the morphotropic boundaries in multiferroics, which are usually identified as monoclinic phase regions by scattering. They offer rich evidence for spatially modulated structures in electron microscopy [27, 28, 29, 30]. In particular the apparent "orthorhombic" phase in $PMN-PbTiO_3$ exists as an adaptive tetragonal phase [27]. Domains with low domain-wall energy corresponding to monoclinic ferroelectric states [28] and pseudo-monoclinic phase [29] were revealed near the morphotropic phase boundaries in $Pb_xZr_{1-x}TiO_3$. Also it is worth mentioning the adaptive phases in shape-memory martensite alloys, which are, in fact, adaptive modulations [30].

**1.1. Flexoelectric effects.** To get insight into the physical properties of domain walls and interfaces in multiferroics at the meso- and nanoscale, deep understanding of *flexo-type couplings* between the gradients of the polar and other order parameters is extremely important. The coupling between the polarization gradient components $\partial P_k / \partial x_l$ and other order parameters contribution to the thermodynamic potential is:

$$\delta U = K_{ijkl} w_{ij} \frac{\partial P_k}{\partial x_l}, \qquad (1)$$

where $K_{ijkl}$ is the corresponding "flexo-type" tensor. Relationships between the dyadic tensor $w_{ij}$ and the order parameters are listed in the **Table 1.**

**Table 1.** Flexo-type coupling in multiferroics allowed by symmetry

| Coupling title | Tensor $w_{ij}$ | Description | Ref. |
|---|---|---|---|
| Flexoelectric | $u_{ij}$ | $u_{ij}$ - the strain tensor | [31, 32] |
| Flexomagnetoelectric | $M_i M_j$ | $M_i$ - spontaneous magnetization | [33, 34] |
| Flexoantimagnetoelectric | $L_i L_j$ | $L_i$ - antiferromagnetic order parameter (e.g. the difference of sub-lattices magnetization) | [35, 36, 37] |
| Flexoferroelectric | $P_i P_j$ | $P_i$ - spontaneous polarization | [38, 39] |
| Flexoantiferroelectric | $A_i A_j$ | $A_i$ - antiferroelectric order parameter (e.g. the difference of sub-lattices polarization) | [24] |
| Flexo-antiferrodistortive | $\Phi_i \Phi_j$ | $\Phi_i$ - antiferrodistortive order parameter (e.g. axial vector of oxygen octahedral rotational modes [40]) | this work |

In its initial form the flexoelectric coupling between the polarization and strain gradient is universal for macro and nanoscale objects [41, 42, 43, 44, 45]. Flexoelectric and all other couplings from the **Table 1** lead to the appearance of improper ferroelectricity in multiferroics with the inhomogeneous spontaneous strain [32], magnetization [33, 34], aniferromagnetic [35, 36] or antiferroelectric order parameter [24] or antiferrodistortions. Here, we explore the antiferrodistortive coupling, since antiferrodistortive modes are virtually present in all the perovskites. Besides the novel flexo-antiferrodistortive coupling can be of purely fundamental interest, we will demonstrate that it can be a source of *incommensurate modulation in multiferroics*.

**1.2. Incommensurate phases (ICP) in multiferroics.** ICP itself as well as the mechanisms of commensurate-incommensurate phase transitions are ones of the most intriguing features of multiferroics [46, 47]. Two well-established mean-field Landau-type approaches of ICP description exist. The first one consider a one-component long-range order parameter assuming that its gradient coupling coefficient in the Landau-Ginzburg-Devonshire (LGD) power expansion is negative and positively-defined higher order derivatives cause the ICP [48, 49, 50]. The second approach, that seems more relevant to the ICP in multiferroics description, considers at least a two-component order parameter with positive gradient coefficients, conventional LGD functional for each component, and biquadratic coupling between the order parameters and Lifshitz invariant [38, 47, 51, 52].

In the letter we show that the increase of the flexo-antiferrodistortive coupling strength firstly leads to commensurate-incommensurate phase transitions, and then to the antiferroelectric-like phase appearance in multiferroics. The scenario seems principally different from the known couplings [47-52] and are in agreement with experiments [24-26].

**2. A universal flexo-antiferrodistortive couplings**

The linear-quadratic coupling between the long-range order parameters, antiferrodistortive octahedral rotations $\Phi_i$ and polarization $P_i$ gradient, allowed by any symmetry, and thus *universal for all antiferrodistortive materials with spatial inhomogeneities*, has the form of nonlinear Lifshitz invariant:

$$U_{P\Phi}[\mathbf{P},\mathbf{\Phi}] = \frac{\xi_{ijkl}^{u,\sigma}}{2}\left(\Phi_i\Phi_j\frac{\partial P_k}{\partial x_l} - P_k\frac{\partial(\Phi_i\Phi_j)}{\partial x_l}\right) \qquad (2a)$$

As universal, the flexo-antiferrodistortive coupling (2a) must be included in the LGD thermodynamic potentials. Below we will regard Helmholtz free energy as $\Phi$-P-u representation and Gibbs potential as $\Phi$-P-$\sigma$ representation (u stands for the strain and $\sigma$ for the stress).

Nonzero components of the novel coupling tensor $\xi_{ijkl}^{u,\sigma}$ can be readily determined from the symmetry theory, e.g. for the m3m parent phase of most perovskites they are $\xi_{1111} = \xi_{2222} = \xi_{3333}$, $\xi_{1122} = \xi_{1133} = \xi_{3322}$, $\xi_{1212} = \xi_{1313} = \xi_{2323}$. Numerical values of $\xi_{ijkl}^{u,\sigma}$ can be calculated from the first principles or measured experimentally.

The relationship $\xi_{ijkl}^u = \xi_{ijkl}^\sigma + f_{ijmn}R_{mnkl}$ is valid, where $R_{mnkl}$ is the rotostriction strain tensor and $f_{ijmn}$ is the flexoelectric stress tensor (see Appendix S1 [53]). Physical origin of the "renormalization" term $f_{ijmn}R_{mnkl}$ is the joint action of the "indirect" flexoelectric and rotostriction coupling, since the rotostriction causes the spontaneous strain with components $u_{mn}^S = R_{mnpq}\Phi_p\Phi_q$ [54]. The relationship $\xi_{ijkl}^u = \xi_{ijkl}^\sigma + f_{ijmn}R_{mnkl}$ explains that the direct coupling (2a) cannot be treated as the simple renormalization of the flexo-roto effect described by the term $f_{ijmn}R_{mnkl}$. However, similar to the flexo-roto coupling [54], the gradient coupling induces the polarization variation $P_i(\mathbf{r}) \propto \xi_{ijkl}^{u,\sigma}\partial(\Phi_j\Phi_k)/\partial x_l$ in regions where the tilt is spatially inhomogeneous (domains walls, surfaces, interfaces).

The bilinear antiferroelectric-antiferrodistortive coupling term between the polarization gradient and tilt components product, allowed by the *any symmetry* of parent phase in ABO$_3$ compounds with an antiferrodistortive mode:

$$U_{A\Phi}[\mathbf{A},\mathbf{\Phi}] = \frac{\zeta_{ijk}}{2}\left(A_i \frac{\partial \Phi_k}{\partial x_j} - \Phi_k \frac{\partial A_i}{\partial x_j}\right) \qquad (2b)$$

Coupling **pseudo-**tensor $\zeta_{ijk}$ non-zero components allowed by the material parent phase symmetry can be determined from the symmetry theory for all point groups. Coupling (2b) is invariant in Φ-P-u and Φ-P-σ representations. Numerical value of the nonzero components $\zeta_{ijk}$ can be defined either from experiment or from the first principle calculations. Polar (or true) vector **A** is the "antipolarization", defined as the difference of polarization in the neighboring equivalent cells *a* and *b*, $\mathbf{A} = (\mathbf{P}_a - \mathbf{P}_b)/2$, axial (or pseudo-) vector $\mathbf{\Phi} = (\mathbf{\Phi}_a - \mathbf{\Phi}_b)/2$ is the structural order parameter, corresponding to the antiferrodistortive rotational modes of oxygen octahedral [40], $\mathbf{\Phi}_a = -\mathbf{\Phi}_b$, considered hereinafter. The oxygen octahedra are regarded rigidly connected within the layers, so they can only rotate as a whole and distortive (Jahn-Teller) modes will be neglected. Transformation laws of pseudo-tensor $\zeta_{ijk}^{u,\sigma}$, true vector **A**, pseudo-vector **Φ** and coordinate derivative $\partial/\partial x_l$ are $\tilde{\zeta}_{ijk} = \det(B) B_{im} B_{jg} B_{ks} \zeta_{mgs}$, $\tilde{A}_i = B_{ip} A_p$, $\tilde{\Phi}_k = \det(B) B_{kf} \Phi_f$. Here the summation is performed over the repeating indexes. **B** is the unitary transformation matrix with components $B_{ij}$ (*i,j* = 1,2,3) representing all the elements of the parent phase point symmetry group with convolution $B_{ij} B_{ik} = \delta_{jk}$ and determinant det(**B**) = ±1. For the case the transformation laws become identity without symbol "tilda", e.g. $\zeta_{ijk} = \det(B) B_{im} B_{jg} B_{ks} \zeta_{mgs}$. Elementary derivation listed in **Appendix S1** proves that the invariant (2a) is indeed invariant with respect to the transformation of parent phase point symmetry group using the elements **B** of the and with respect to the permutation operation $a \leftrightarrow b$, when $\mathbf{A} \leftrightarrow -\mathbf{A}$ simultaneously with $\mathbf{\Phi} \leftrightarrow -\mathbf{\Phi}$. For the case of cubic symmetry group m3m we calculated that $\zeta_{ijk}^{u,\sigma} \equiv \chi e_{ijk}$, where symbol $e_{ijk}$ is the antisymmetric Levi-Chivita psevdo-tensor and constant $\chi$ is a true scalar (in particular nonzero components are $\zeta_{123}^{u,\sigma} = -\zeta_{213}^{u,\sigma} = \zeta_{231}^{u,\sigma} = ... \equiv \chi$ since $e_{123} = -e_{213} = e_{231} = ... = 1$). Using the definition of curl operation, for m3m parent phase Eq.(2b) acquires the form $U_{A\Phi}[\mathbf{A},\mathbf{\Phi}] = \frac{\chi}{2}(\mathbf{A} \cdot \text{rot}\mathbf{\Phi} - \mathbf{\Phi} \cdot \text{rot}\mathbf{A})$.

Due to the universality, the flexo-antiferrodistortive coupling (2) can be a significant driving force for the spontaneous onset of spatial modulation in a wide class of partially clamped multiferroics with antiferrodistortive structural order parameter, such as thin films, twin walls and antiphase boundaries in $Eu_xSr_yBa_{1-x-y}TiO_3$, $Bi_ySm_{1-y}FeO_3$, $Sr_yCa_{1-y}TiO_3$, $Sr_yMn_{1-y}TiO_3$, etc.

The large difference between the coupling (2a) and (2b) is in the different tilt power of the bilinear coupling tensors $\zeta_{ijk}$ and $\xi_{ijkl}$ in typical antiferrodistortive-perovskites with m3m parent phase (such as $(Pb,Zr)TiO_3$, $(Sr,Eu)TiO_3$, $CaTiO_3$ and $BiFeO_3$) are listed in **Table 2**. Peculiarities of the bilinear coupling tensors in 32 symmetry classes are listed in **Table 3**.

**Table 2.** Symmetry of the bilinear coupling in typical antiferrodistortive-perovskites

| Point group symmetry and $ABO_3$ example | Tensor symbol, structure and/or nontrivial components | |
|---|---|---|
| | $\zeta_{ijk}$ in Eq.(2a) | $\xi_{ijkl}$ in Eq.(2b) |
| m3m parent phase of most perovskites | $\zeta_{123} = -\zeta_{213} = \zeta_{231} = -\zeta_{132} = \zeta_{312} = -\zeta_{321}$ $\zeta_{ijk} \equiv \chi e_{ijk}$, $\chi$ is a scalar | $\xi_{1111} = \xi_{2222} = \xi_{3333}$, $\xi_{1122} = \xi_{1133} = \xi_{3322}$, $\xi_{1212} = \xi_{1313} = \xi_{2323}$ |
| 4mm $(Pb,Zr)TiO_3$ | $\zeta_{123} = -\zeta_{213}$, $\zeta_{312} = -\zeta_{321}$, $\zeta_{132} = -\zeta_{231}$ | $\xi_{1111}$, $\xi_{1122}$, $\xi_{1212}$, $\xi_{1133}$, $\xi_{3311}$, $\xi_{1313}$, $\xi_{1331}$, $\xi_{3333}$ |
| 4/mmm $(Sr,Eu)TiO_3$ | $\zeta_{123} = -\zeta_{213}$, $\zeta_{312} = -\zeta_{321}$, $\zeta_{132} = -\zeta_{231}$ | Ibidem to 4mm |
| mmm $CaTiO_3$ | $\zeta_{123}$, $\zeta_{132}$, $\zeta_{312}$, $\zeta_{213}$, $\zeta_{231}$, $\zeta_{321}$ (all are different) | $\xi_{1111}$, $\xi_{1122}$, $\xi_{2211}$, $\xi_{2222}$, $\xi_{1212}$, $\xi_{1221}$, $\xi_{2233}$, $\xi_{1133}$, $\xi_{3311}$, $\xi_{1313}$, $\xi_{1331}$, $\xi_{2323}$, $\xi_{2332}$, $\xi_{3333}$ |
| mm2 $CaTiO_3$ | $\zeta_{123}$, $\zeta_{132}$, $\zeta_{312}$, $\zeta_{213}$, $\zeta_{231}$, $\zeta_{321}$ (all are different) | Ibidem to mmm |
| 3m $BiFeO_3$ | $\zeta_{112} = \zeta_{121} = \zeta_{211} = -\zeta_{222}$, $\zeta_{123} = -\zeta_{213}$, $\zeta_{312} = -\zeta_{321}$, $\zeta_{132} = -\zeta_{231}$ | $\xi_{1111}$, $\xi_{1122}$, $\xi_{1133}$, $\xi_{1313}$, $\xi_{1331}$, $\xi_{3311}$, $\xi_{3333}$, $\xi_{1113}$, $\xi_{1131}$, $\xi_{1311}$ |

**Table 3.** Peculiarities of the bilinear coupling tensors in 32 symmetry classes

| | | Tensor symbol, structure and/or nontrivial components | | | | | |
|---|---|---|---|---|---|---|---|
| | | $\zeta_{ijk}$ in Eq.(2a) | | | $\xi_{ijkl}$ in Eq.(2b) | | |
| | symmetry class | Non-zero comp. | Different comp. | Equal in module but different in sign comp. | Nonzero comp. | Different comp. | Equal in module but different in sign comp. |
| 1. | 1 | 27 | 27 | - | 81 | 54 | - |
| 2. | $\bar{1}$ | 27 | 27 | - | 81 | 54 | - |
| 3. | 2 | 13 | 13 | - | 41 | 28 | - |
| 4. | m | 13 | 13 | - | 41 | 28 | - |
| 5. | $2/m$ | 13 | 13 | - | 41 | 28 | - |
| 6. | 222 | 6 | 6 | - | 21 | 15 | - |
| 7. | mm2 | 6 | 6 | - | 21 | 15 | - |
| 8. | *mmm* | 6 | 6 | - | 21 | 15 | - |
| 9. | 4 | 13 | 10 | 3 | 39 | 20 | 6 |
| 10. | $\bar{4}$ | 13 | 10 | 3 | 39 | 20 | 6 |
| 11. | $\bar{4}2m$ | 6 | 6 | 3 | 21 | 9 | - |
| 12. | 422 | 6 | 6 | 3 | 21 | 9 | - |
| 13. | $4/m$ | 13 | 10 | 3 | 39 | 20 | 6 |
| 14. | $4mm$ | 6 | 6 | 3 | 21 | 9 | - |
| 15. | $4/mmm$ | 6 | 6 | 3 | 21 | 9 | - |
| 16. | 3 | 21 | 14 | 5 | 71 | 36 | 11 |
| 17. | 32 | 10 | 8 | 4 | 38 | 16 | 3 |
| 18. | $3m$ | 10 | 8 | 4 | 38 | 16 | 3 |
| 19. | $\bar{3}$ | 21 | 14 | 5 | 71 | 36 | 11 |
| 20. | $\bar{3}m$ | 6 | 6 | 3 | 38 | 16 | 3 |
| 21. | $\bar{6}$ | 13 | 10 | 3 | 39 | 20 | 6 |
| 22. | $\bar{6}m2$ | 6 | 6 | 3 | 21 | 8 | - |
| 23. | 6 | 13 | 10 | 3 | 39 | 20 | 6 |
| 24. | 622 | 6 | 6 | 3 | 21 | 8 | - |
| 25. | $6/m$ | 13 | 10 | 3 | 39 | 20 | 6 |
| 26. | $6mm$ | 6 | 6 | 3 | 21 | 8 | - |
| 27. | $6/mmm$ | 6 | 6 | 3 | 21 | 8 | - |
| 28. | 23 | 6 | 2 | - | 21 | 5 | - |
| 29. | $m3$ | 6 | 2 | - | 21 | 5 | - |
| 30. | $\bar{4}3m$ | 6 | 2 | 1 | 21 | 3 | - |
| 31. | 432 | 6 | 2 | 1 | 21 | 3 | - |
| 32. | $m3m$ | 6 | 2 | 1 | 21 | 3 | - |

### 3. Illustration of seeming paradox on twin walls in ferroelastics

For multiferroics with antiferrodistortive and polar long-range order parameters the conventional form of the bulk LGD Helmholtz and Gibbs functional density are: $F_b[\mathbf{P},\mathbf{\Phi},\mathbf{u}] = U_{LGD}^u + U_{Elastic}^u + U_{P\Phi}^u + U_{A\Phi}^u$ and $G_b[\mathbf{P},\mathbf{\Phi},\mathbf{\sigma}] = U_{LGD}^\sigma + U_{Elastic}^\sigma + U_{P\Phi}^\sigma + U_{A\Phi}^\sigma$. A typical form of the LGD contribution $U_{LGD}^{u;\sigma}$ as a function of the octahedral rotations $\Phi_i$ and polarization $P_i$ and elastic contribution $U_{Elastic}^{u;\sigma}$ that includes purely elastic, electrostriction, rotostriction and flexoelectric coupling terms are listed in the Appendix S1 of [53]. The flexo-antiferrodistortive coupling terms $U_{P\Phi}^{u;\sigma}$ and $U_{A\Phi}$ are given by Eq.(2). Below we will regard $F_b[\mathbf{P},\mathbf{\Phi},\mathbf{u}]$ as Φ-P-u representation and $G_b[\mathbf{P},\mathbf{\Phi},\mathbf{\sigma}]$ as Φ-P-σ representation. Starting from the variation of the functional in any of representations, Euler-Lagrange equations for the polarization and tilt as well as equations of state for the elastic stress or strain can be derived. The equations of state give the relation between the stress and strain. After the substitution of the relation into the Euler-Lagrange equations, unambiguous relationship between the coefficients of LGD expansion for Φ-P-u representation and Φ-P-σ representation can be established. In particular the biquadratic flexo-antiferrodistortive coupling tensor (2a) transforms as $\xi_{ijkl}^{u} = \xi_{ijkl}^{\sigma} + f_{ijmn}R_{mnkl}$, and so if one starts from conventional Φ-P-σ representation with zero $\xi_{ijkl}^{\sigma}$, then mandatory come to nonzero values $\xi_{ijkl}^{u} = f_{ijmn}R_{mnkl}$ in Φ-P-u representation. In other words the condition $\xi_{ijkl}^{u} = \xi_{ijkl}^{\sigma} = 0$ never can be valid and its artificial fulfillment can lead to unphysical paradox.

For demonstration of the paradox appeared when the universal bilinear coupling (2a) *is not included properly* into the LGD potential, we chose ferroelastic SrTiO$_3$, because all its material parameters are relatively well-known, including rotostriction [55, 56, 57], and the flexoelectric coupling tensor components were measured experimentally [58, 59, 60],. SrTiO$_3$ undergoes the second order phase transition at $T \approx 105$ K from cubic phase of m3m symmetry to tetragonal antiferrodistortive phase of 4/mmm symmetry with one-component spontaneous tilt $\Phi_S$. Single-domain regions of bulk SrTiO$_3$ are non-polar, while the flexo-roto coupling can induce a spontaneous polarization in the vicinity of elastic domain walls [54]. Let us consider the typical head-to-head and head-to-tail twin boundaries (TB) between domains "1" and "2" with different orientation of tilts very far from the wall (see **Fig. 1a**). **Figure 1b** shows the distribution of the tilts $\widetilde{\Phi}_1 \perp$ TB and $\widetilde{\Phi}_2 \uparrow\uparrow$ TB calculated across head-to-head TB in Φ-P-σ (solid curves) and Φ-P-u (dash-dotted curves) representations without the coupling term (2). The minute difference between the solid and dash-dotted curves originated from the flexoelectric coupling. Polarization components $\widetilde{P}_1 \perp$ TB and $\widetilde{P}_2 \uparrow\uparrow$ TB are induced by the flexo-coupling across the TB and vanish far from it for both Φ-P-σ (solid curves) and Φ-P-u (dash-dotted curves) representations, but corresponding curves in **Fig. 1c-d** look very different. Moreover, $\widetilde{P}_2$ is absent in Φ-P-σ representation at temperatures higher about 36 K, but rather high in Φ-P-u representation (**Fig. 1d**). *Again, the curves for $\widetilde{P}_1$ and $\widetilde{P}_2$ are calculated without the coupling term (2a).*

So, one can see the *seeming paradox here*. It is well-known that Gibbs and Helmholtz functionals are different, but the values calculated from the Euler-Lagrange equations should be the same. So, what is the physical origin of the difference in polarization profiles calculated in Φ-P-σ and Φ-P-u representations? The additional "roto-flexo" sources of polarization, $\widetilde{f}_{66}\widetilde{R}_{66}\partial(\widetilde{\Phi}_1\widetilde{\Phi}_2)/\partial\widetilde{x}_1$, appeared in the Euler-Lagrange equations in Φ-P-u representation in contrast to Φ-P-σ one (see Appendix S2-3 [53]). By the addition of the coupling (2) the term renormalizes as $(\widetilde{\xi}_{66}^{\sigma} + \widetilde{f}_{66}\widetilde{R}_{66})\partial(\widetilde{\Phi}_1\widetilde{\Phi}_2)/\partial\widetilde{x}_1$ and becomes zero for $\widetilde{\xi}_{66}^{\sigma} = -\widetilde{f}_{66}\widetilde{R}_{66}$. If one starts from the conventional Φ-P-σ representation with zero $\xi_{ijkl}^{\sigma} \equiv 0$, that corresponds to solid and dotted curves in **Fig. 1c,d**, then mandatory $\xi_{ijkl}^{u} \equiv f_{ijmn}R_{mnkl}$, but not $\xi_{ijkl}^{u} \equiv 0$ as regarded for dashed and dash-dotted **curves.** These results allow one to regard the value $\xi^u_{ijkl} \propto f_{ijmn} R_{mnkl}$ as a reasonable estimation of the flexo-antiferrodistortive coupling strength in SrTiO$_3$. This gives $\xi^u_{11} \propto -5.08 \times 10^{19}$, $\xi^u_{12} \propto 2.66 \times 10^{19}$ and $\xi^u_{11} \propto -1.95 \times 10^{19}$ V/m$^2$.

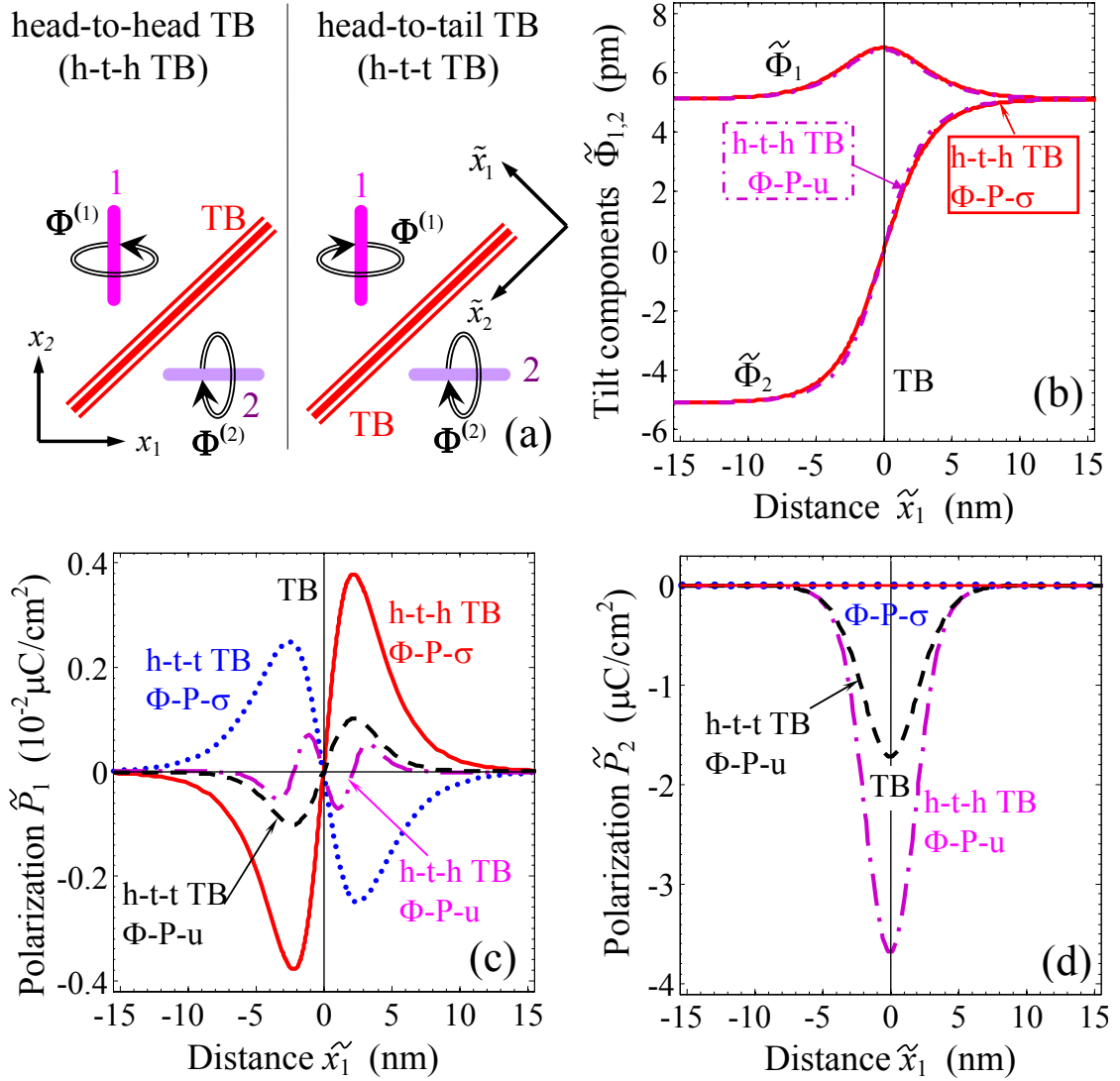

**Figure 1.** (a) Schematics of head-to-head (h-t-h) and head-to-tail (h-t-t) twins $\mathbf{\Phi}^{(1)} = (\pm \Phi_S, 0, 0)$ and $\mathbf{\Phi}^{(2)} = (0, \Phi_S, 0)$, where the signs "$\pm$" correspond to the orientation of the tilt arrows far from the TB. Rotated coordinate system is $\{\tilde{x}_1, \tilde{x}_2\}$. Profiles of tilts $\tilde{\Phi}_1(\tilde{x}_1)$ and $\tilde{\Phi}_2(\tilde{x}_1)$ (b) and polarization components $\tilde{P}_1(\tilde{x}_1)$ (c) and $\tilde{P}_2(\tilde{x}_1)$ (d) across the TB in SrTiO$_3$ at temperature 50 K. Different curves were calculated using Φ-P-σ representation with $\xi^\sigma_{ijkl} = 0$ and Φ-P-u representation with $\xi^u_{ijkl} = 0$. \

Beyond the paradox resolution for the SrTiO$_3$ example, we predicted a noticeable influence of the flexo-antiferrodistortive coupling on the structure and physical properties of the of domain boundaries in the related $Eu_xSr_yBa_{1-x-y}TiO_3$ systems. Significant differences between Φ-P-u and Φ-P-σ curves in **Fig. 1c,d** give all grounds to expect that the influence can be rather strong and thus flexo-antiferrodistortive coupling should be mandatory to be taken into account in future.

### 4. Modulated phases caused by the flexo-antiferrodistortive coupling

Here we illustrate that the flexo-antiferrodistortive coupling strongly influences the order parameters, phase stability regions of multiferroics and leads to the appearance of the incommensurately modulated phases (MP). In order to derive analytical results, let us consider one-component tilt $\Phi(x)$ and polarization $P_3(x_1) \equiv P(x)$ (1D theory without depolarization effects). Along with the coupling (2a) LGD potential density acquires the form:

$$F_b[P,\Phi] = \begin{pmatrix} \beta_1\Phi^2 + \beta_{11}\Phi^4 + \dfrac{v}{2}\left(\dfrac{\partial\Phi}{\partial x}\right)^2 + \alpha_1 P^2 + \alpha_{11}P^4 + \dfrac{g}{2}\left(\dfrac{\partial P}{\partial x}\right)^2 \\ -\eta P^2\Phi^2 + \dfrac{\xi}{2}\left(\dfrac{\partial P}{\partial x}\Phi^2 - P\dfrac{\partial(\Phi^2)}{\partial x}\right) \end{pmatrix} \quad (3)$$

Coefficients $\alpha_1$ and $\beta_1$ are linear inverse susceptibilities of $P$ and tilt $\Phi$; $\alpha_{11}$ and $\beta_{11}$ are nonlinear generalized stiffness for corresponding order parameter; tilt and polarization gradient coefficients are $v$ and $g$. The flexo-antiferrodistortive coupling coefficient is $\xi$; $\eta$ is the biquadratic coupling coefficient. For most of multiferroics and their solid solutions linear dependences of the coefficients $\alpha_1$ and $\beta_1$ on temperature $T$ are valid, $\alpha_1(T) = \alpha_T(T - T_P)$ and $\beta_1(T) = \beta_T(T - T_\Phi)$, where the polar and antiferrodistortive critical temperatures $T_P$ and $T_\Phi$ can be dependent on the chemical composition of multiferroic solid solution. Other coefficients are typically weakly (or at least non-critically) temperature dependent, but can be strongly composition dependent. The energy (3) is stable at high values of order parameters under the conditions $\alpha_{11} > 0$, $\beta_{11} > 0$, $2\sqrt{\alpha_{11}\beta_{11}} - \eta > 0$, $v > 0$ and $g > 0$. Thermodynamically stable phases described by the energy (3), corresponding order parameters values and stability conditions are listed in the **Table 4.**

**Table 4.** Thermodynamically stable bulk phases of the free energy (3)

| Phase description and abbreviation | Order parameters | Stability condition |
|---|---|---|
| Parent (**PP**) | $P = \Phi = 0$ | $\alpha_1 > 0$, $\beta_1 > 0$ |
| Antiferrodistortive (normal **AFD**) | $\Phi = \pm\sqrt{-\beta_1/2\beta_{11}}$, $P = 0$ | $\beta_1 < 0$, $2\alpha_1 + \eta(\beta_1/\beta_{11}) > 0$ |
| Antiferrodistortive-ferroelectric (normal **AFD+ FE**) | $\Phi = \sqrt{-\dfrac{2\beta_1 + \eta(\alpha_1/\alpha_{11})}{4\beta_{11} - (\eta^2/\alpha_{11})}}$, | $2\alpha_1 + \eta(\beta_1/\beta_{11}) < 0$, $2\beta_1 + \eta(\alpha_1/\alpha_{11}) < 0$, $2\sqrt{\alpha_{11}\beta_{11}} > -\eta$ |
| | $P = \sqrt{-\dfrac{2\alpha_1 + \eta(\beta_1/\beta_{11})}{4\alpha_{11} - (\eta^2/\beta_{11})}}$ | |
| Ferroelectric (**FE**) | $P = \sqrt{-\alpha_1/2\alpha_{11}}$, $\Phi = 0$ | $\alpha_1 < 0$, $2\beta_1 + \eta(\alpha_1/\alpha_{11}) > 0$ |
| Incommensurate AFD modulated phase (**MP**) with possible **AFE** phase | In harmonic approximation $P = P_0 - \delta P \sin(kx)$, $\Phi = \Phi_0 - \delta\Phi\cos(kx)$ | $\xi^2 \geq \beta_{11}\left(\sqrt{2g} + \sqrt{-\dfrac{v}{2\beta_1}\left(2\alpha_1 + \eta\dfrac{\beta_1}{\beta_{11}}\right)}\right)^2$ $2\alpha_1 + \eta(\beta_1/\beta_{11}) > 0$, $\beta_1 < 0$ |

Assuming temperature dependencies $\alpha_1(T) = \alpha_T(T - T_P)$ and $\beta_1(T) = \beta_T(T - T_\Phi)$ in the functional (3), the system phase diagram depends on the 5 dimensionless parameters, namely flexo-antidistortive and biquadratic coupling constants, $\xi^* = |\xi|/\left(2\sqrt{v\sqrt{\alpha_{11}\beta_{11}}}\right)$ and $\eta^* = \eta/\sqrt{\alpha_{11}\beta_{11}}$, temperature $(T - T_\Phi)/(T_\Phi - T_P) = t$, ratios $\Delta = \beta_T\sqrt{\alpha_{11}}/(\alpha_T\sqrt{\beta_{11}})$ and $g^* = g\sqrt{\alpha_{11}}/(v\sqrt{\beta_{11}})$. Appeared that the values $\xi^*$ and $\eta^*$ define the phase diagram. Temperature *t* should be negative in the ordered phase, its value determine the position of the vertical boundary between AFD+FE, AFD and FE phases. Oxygen octahedrons arrangement in AFD, AFD+FE and MP phases is schematically shown in the **Figure 2a**. **Figure 2b** illustrates the typical evolution of the phases in dependence on the $\xi^*$ and $\eta^*$. MP region is not very sensitive to the values of $\eta^*$ and *t*, but requires $|\xi^*|$ values higher than 2. MP strongly enlarges the stability region with the $|\xi^*|$ increase. AFD, AFD+FE and FE phase boundaries appeared indeed sensitive to the values of $\eta^*$ and *t*.

To study analytically the modulation period, that is the most important feature of the MP, we use the harmonic modulation approximations (HMA) for the order parameters distributions:

$$P = P_0 - \delta P\sin(kx), \qquad \Phi = \Phi_0 - \delta\Phi\cos(kx). \tag{4a}$$

HMA is valid in the vicinity of the MP boundaries. The modulation number *k*, "base" $\Phi_0$ and $P_0$, amplitudes $\delta P$ and $\delta\Phi$ are variational parameters determined from the energy (3) minimization (see Appendix S4 [53]). The relationship between *k* and $\xi$ is:

$$\dfrac{vg}{4k^2\Phi_0^2}\left(k^2 + \dfrac{1}{L_\Phi^2}\right)\left(k^2 + \dfrac{1}{L_P^2}\right) = \xi^2. \tag{4b}$$

The solution of biquadratic Eq.(4b) for the modulation vector is $k_\pm = \sqrt{\left(-b \pm \sqrt{b^2 - 4c}\right)/2}$, where $b = L_P^{-2} + L_\Phi^{-2} - 4\xi^2\Phi_0^2/(vg)$ and $c = L_P^{-2}L_\Phi^{-2}$. Here we introduced the polar and structural correlation lengths as $L_P = \sqrt{g(2\alpha_1 + \eta(\beta_1/\beta_{11}))^{-1}}$ and $L_\Phi = \sqrt{v/(-4\beta_1)}$ correspondingly, which are positive in the AFD MP.

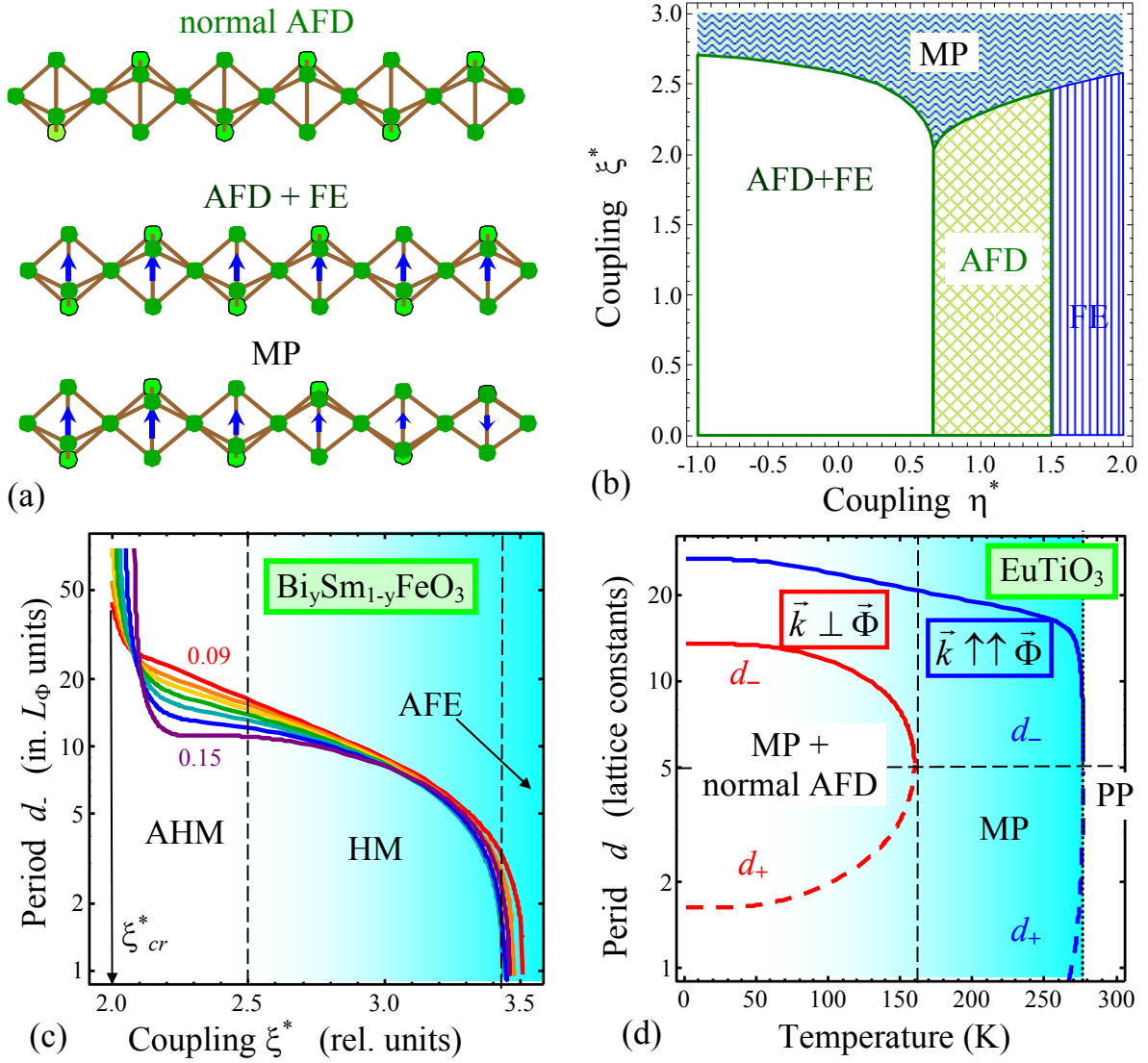

**Figure 2. (a)** Oxygen octahedrons arrangement in AFD, AFD+FE and MP phases. Blue arrows indicate the polarization. **(b)** Phases evolution in dependence on the coupling constants $\xi^*$ and $\eta^*$ calculated for $\Delta = 0.5$, $g^* = 8$ and $t = -1.5$. **(c)** IC AFD-modulation period $d/L_\Phi$ vs. the coupling strength $\xi^*$ calculated for $Bi_ySm_{1-y}FeO_3$ with Sm content $y = 0.09 – 0.15$ (different curves with y-step of 0.01) and $T = 300$ K. AHM indicates the region with anharmonic modulation, HM indicates the harmonic modulation, AFE is the antiferroelectric-like region. **(d)** Temperature dependence of the IC AFD-modulation periods $d_\pm$ calculated in the MP of $EuTiO_3$ for 2 modulation directions $\vec{k}_\pm \uparrow\uparrow \vec{\Phi}$ and $\vec{k}_\pm \perp \vec{\Phi}$. $EuTiO_3$ parameters are listed in the text.

Using that $|\Phi_0| \gg |\delta\Phi|$, $\Phi_0 \approx \pm\sqrt{-\beta_1/2\beta_{11}}$, $P_0 = 0$ and $\delta P$ is small in the vicinity of MP-AFD boundary, approximate expressions for the MP-AFD phase boundary and the wave vector at the boundary $k_b$ were derived:

$$\xi^2 = \frac{g\nu}{4\Phi_0^2}\left(\frac{1}{L_P}+\frac{1}{L_\Phi}\right)^2, \qquad k_b = \sqrt{\frac{1}{L_P L_\Phi}}. \tag{5}$$

In fact Eq.(5) determines the minimal critical value of the coupling strength, $\xi_{cr}$, required for the commensurate-incommensurate phase transition. The physical sense of the condition $|\xi| \geq \xi_{cr}$ is that the effective length induced by the flexo-antiferrodistortive coupling should be higher than the sum of inverse polar and structural correlation lengths, since exactly $\xi_{cr}^2 \propto (L_P^{-1}+L_\Phi^{-1})^2$ per Eq.(5). The modulation profile is quasi-harmonic if the half-period $\pi/k$ is not much higher than the effective correlation length, $L_C = L_P L_\Phi/(L_P+L_\Phi)$. Though the main results (2)-(5), **Tables 2-4** and phase evolution shown in **Fig.2b** are *universal* (i.e. not material-specific), let us consider briefly their applications for the determination of the IC modulation period in concrete materials.

IC MP was observed in antiferrodistortive multiferroic $Bi_y Sm_{1-y} FeO_3$ for Sm content $y \propto 0.1$ at room temperature [24]. Using Eqs.(3)-(4b) for $Bi_y Sm_{1-y} FeO_3$ parameters, namely the transition temperature from cubic phase into the orthorhombic one, $T_\Phi(y) = T_{\Phi 0} + (T_{\Phi 1} - T_{\Phi 0})y$, $T_{\Phi 0} = 1200$ K and $T_{\Phi 1} = 1100$ K [61], ferroelectric Curie temperature $T_P(y) = T_{P0}(1-(y/y_{cr}))^{1/2}$, $T_{P0} = 1120$ K and $y_{cr} = 0.16$ [62], we calculated the dependence of the modulation period $d_\pm = 2\pi/k_\pm$ on the flexo-antiferrodistortive coupling value $\xi^*$. **Figure 2c** illustrates that the IC modulation period $d_-/L_\Phi$ diverges at $\xi^* \to \xi_{cr}^*$, then rapidly decreases with $\xi^*$ increase, and becomes zero at $\xi^* \approx 3.5$. When the period becomes compatible or smaller than the lattice constant, it indicates the origin of antiferroelectric-like (AFE) polarization. Since typically $L_C$ is about a lattice constant, the inequality $d \leq L_C$ determines the AFE phase transition that appeared at $\xi^* \geq 3.4$ in agreement with experiment [24].

Goian et al [25] observed the incommensurate AFD tetragonal structure in $EuTiO_3$ with periodicity of about 16 *unit cells* below 300 K. Kim and Ryan [26] reported about the incommensurate AFD-AFE superstructure periodicity of about *14 unit cells* below 285 K in $EuTiO_3$. Using Eqs.(3)-(4b) for $EuTiO_3$, we assume that coefficient $\alpha_1(T)$ depends on temperature $T$ in accordance with Barrett law [63], $\alpha_1(T) = \alpha_T^{(P)} T_q^{(P)} (\coth(T_q^{(P)}/T) - \coth(T_q^{(P)}/T_c^{(P)}))$. Coefficient $\alpha_T^{(P)} = 0.98 \times 10^6$ m/(F K), $T_q^{(P)} \approx 115$ K is the called quantum vibration temperature, $T_c^{(P)} \approx -133$ K is the "effective" Curie temperature corresponding to polar soft mode in bulk $EuTiO_3$ [64, 65]. To account for the experiment and Barrett law the coefficient $\beta_1(T)$ depends on temperature as $\beta_1(T) = \beta_T^{(\Phi)} T_q^{(\Phi)} (\coth(T_q^{(\Phi)}/T) - \coth(T_q^{(\Phi)}/T_c^{(\Phi)}))$, where $\beta_T^{(\Phi)} = 1.96 \times 10^{26}$ J/(m$^5$ K), $T_q^{(\Phi)} \approx 102$ K, $T_c^{(\Phi)} \approx 281$ K [66, 67]. In particular this gives that $\beta_1 = -3.75 \times 10^{28}$ J/m$^5$ and $\alpha_1 = 2.73 \times 10^8 10^6$ m/F at $T$=5.2 K. Parameters $\beta_{11} = 0.436 \times 10^{50}$ J/m$^7$, $v_{11} = 0.28 \times 10^{10}$ J/m$^3$, $v_{44} = 7.34 \times 10^{10}$ J/m$^3$, $\alpha_{11} = 1.6 \times 10^9$ m$^5$/(C$^2$F), $g \approx 0.3 \times 10^{-10} \times$V·m$^3$/C, $\eta_{11} = 2.23 \times 10^{29}$(F m)$^{-1}$, $\eta_{12} = -0.85 \times 10^{29}$(F m)$^{-1}$ [53] and modulation period $d \cong (10-20)$ lattice constants (*l.c.*) we estimated the coupling constant $\xi$. Depending on the tilt orientation $\vec{\Phi}$ with respect to the direction of modulation vector $\vec{k} \uparrow\uparrow x$ we obtained that $\xi = \xi_{11} = 3 \times 10^{20}$ V/m$^2$ for the case $\vec{k} \uparrow\uparrow \vec{\Phi}$ when $v = v_{11}$ and $\eta = \eta_{11}$, meanwhile $\xi = \xi_{12} = 2.6 \times 10^{20}$ V/m$^2$ for the case $\vec{k} \perp \vec{\Phi}$, when $v = v_{44}$ and $\eta = \eta_{12}$. Shown in **Figure 2d** IC modulation periods $d_\pm = 2\pi/k_\pm$ are with the range 2-20 *l.c.* depending on temperature and $\vec{\Phi}$ orientation. When the period $d_+$ becomes compatible or smaller than 2 *l.c.* for $\vec{k}_+ \uparrow\uparrow \vec{\Phi}$ and temperatures lower 110 K, it may indicate the origin of coexisted AFE polarization.

Emergence of spatially modulated polarization and tilt at the structural domain walls of EuTiO$_3$ are shown in the **Figure 3.** The modulation is absent without the flexoantiferrodistortive coupling as well as when the coupling strength is smaller that the critical value (see solid curves). It originates for the coupling strength higher then the critical value dependent on the modulation vector $\vec{k}$ orientation with respect to the tilt $\vec{\Phi}$ (compare dashed curves for $\vec{k} \uparrow\uparrow \vec{\Phi}$ and $\vec{k} \perp \vec{\Phi}$). Finally the polarization modulation amplitude increases and its period decreases with the coupling value increase (see dotted curves which look completely incommensurate). Note, that in agreement with experimental observations [26] the spatial modulation of polarization can acquire antiferroelectric features for the case $\vec{k} \perp \vec{\Phi}$ and high values of $\xi = 3 \times 10^{20}$ V/m$^2$ (see dotted curves in **Fig.3b**).

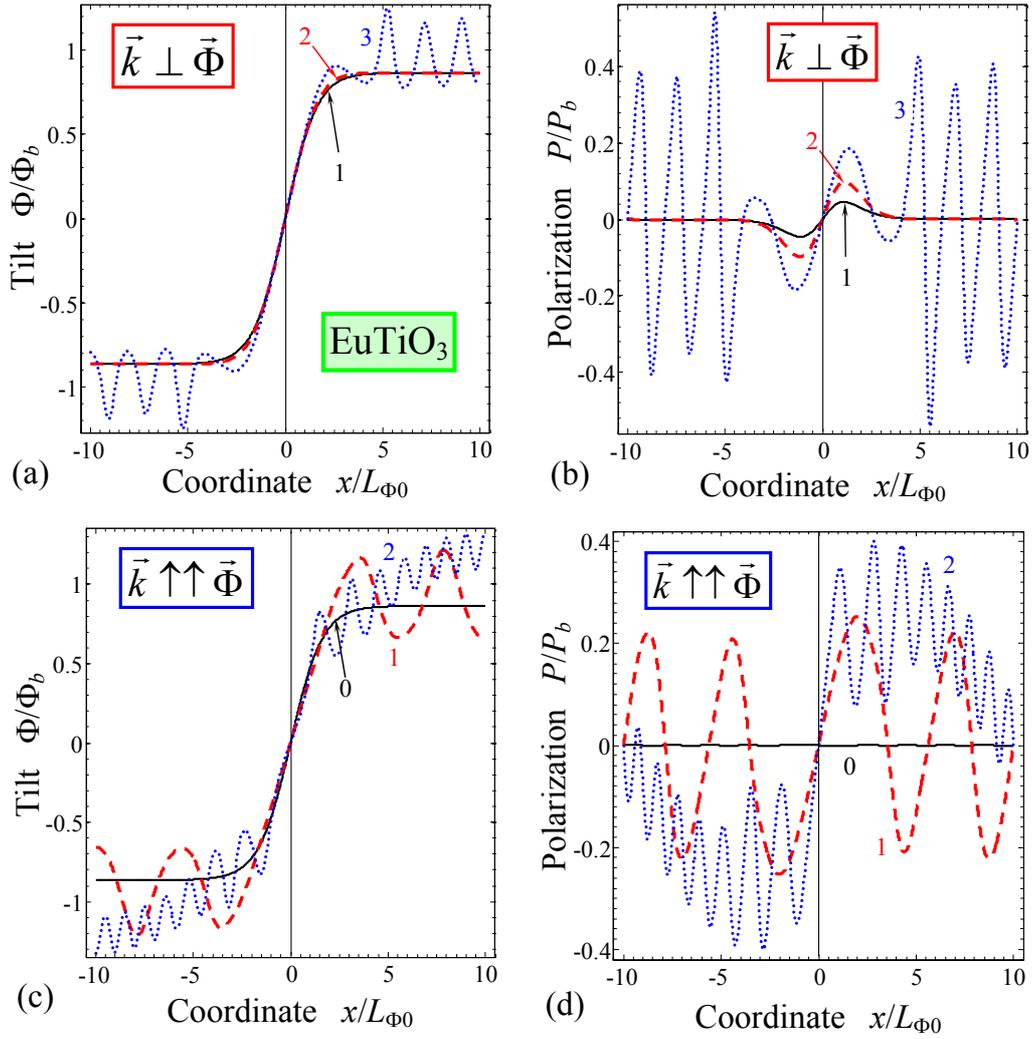

**Figure 3.** Tilt **(a,c)** and polarization **(b,d)** spatial modulation originated near EuTiO$_3$ antiphase boundary between the domains with opposite orientations of tilt vector calculated for different values of flexoantiferrodistortive coupling coefficient. For **orientation** $\vec{k} \perp \vec{\Phi}$ **(a,b)** the value $\xi = \xi_{12} = 1$, $2$ and $3 \times 10^{20}$ V/m$^2$ (solid, dashed and dotted curves respectively). For **orientation** $\vec{k} \uparrow\uparrow \vec{\Phi}$ **(c,d)** the value $\xi = \xi_{11} = 0$, $1$ and $2 \times 10^{20}$ V/m$^2$ (solid, dashed and dotted curves respectively). Temperature $T$=200 K, scales for the tilt, polarization and x-coordinate are introduced as $\Phi_b = \sqrt{\beta_T^{(\Phi)} T_q^{(\Phi)} / (2\beta_{11})}$, $P_b = \sqrt{\beta_T^{(\Phi)} T_q^{(\Phi)} / 2\sqrt{\beta_{11}\alpha_{11}}}$ and $L_{\Phi 0} = \sqrt{v/(2\beta_T^{(\Phi)} T_q^{(\Phi)})}$, where $v = v_{11}$ for $\vec{k} \uparrow\uparrow \vec{\Phi}$ and $v = v_{44}$ for $\vec{k} \perp \vec{\Phi}$. EuTiO$_3$ parameters are listed in the text.

## 5. Summary

Our analysis provides new insight into origins of morphotropic and nanoscale phase-separated systems in complex oxide multiferroics that have eluded macroscopic description. Namely, we show that a universal flexo-antiferrodistortive coupling between the (anti)polarization gradient and the structural long-range order parameter should be included in the LGD functional of antiferrodistortive materials. The coupling tensor components should be either calculated from the first principles or determined experimentally. Using the classical example of ferroelastic twin walls in antiferrodistortive incipient ferroelectric SrTiO$_3$ we estimated the coupling strength as the convolution of the flexoelectric and rotostriction coupling tensors.

The coupling strongly influences on the physical properties and phase diagrams of the multiferroic with ferroelectric and antiferrodistortive phases, primary leading to the appearance of the spatially modulated mixed phases. Incommensurate modulation appears spontaneously when the coupling strength exceeds the critical value proportional to the sum of inverse polar and structural correlation lengths (commensurate-incommensurate phase transition). We demonstrated that modulated phase appears and strongly enlarges its stability region with the coupling strength increase. Further increase of the coupling strength can lead to the modulated antiferroelectric-like antiferrodistortive phase. The scenario seems principally different from the known ones [47-52] and universal.

Concrete examples of the proposed scenario applicability to real systems, where the incommensurate modulation was observed, are multiferroics solid solution Bi$_y$Sm$_{1-y}$FeO$_3$ [24] and single-phase EuTiO$_3$ [25, 26]. Also the proposed description could be helpful in the design of the ferroics with advanced properties. A promising candidate could be a Eu$_x$Sr$_y$Ba$_{1-x-y}$TiO$_3$ solid solution [68], recently used as a successful alternative to Eu$_x$Ba$_{1-x}$TiO$_3$ ceramics for search of the fundamental electric dipole moment of the electron [69, 70, 71]. Our approach establishes the ranges of possible phases as a function of composition and temperature and thus can help to design magnetoelectric Eu$_x$Sr$_y$Ba$_{1-x-y}$TiO$_3$ solid solutions, which have purely ferroelectric phase without incommensurate antiferrodistortive and/or modulated phases, where the high-order magnetoelectric coupling is suppressed.

**Acknowledgements**

The work was supported in part (SVK, AB) by the US Department of Energy, Basic Energy Sciences, Materials Sciences and Engineering Division, the national science foundation DMR-1210588 and DMR-0820404. (VG, LQC, and YG) ANM and EAE acknowledge the State Fund of Fundamental State Fund of Fundamental Research of Ukraine, SFFR-NSF project UU48/002 (NSF number DMR-1210588).

# Supplemental Materials
# Appendix S1
## S1.1. Invariant form of the flexo-antiferrodistortive couplings

For multiferroics with the antiferrodistortive and polar long-range order parameters the conventional form of the bulk LGD Helmholtz free energy and Gibbs potential density are:

$$F_b[\mathbf{P},\mathbf{\Phi},\mathbf{u}] = U^u_{LGD} + U^u_{Elastic} + U^u_{P\Phi} + U_{A\Phi}, \quad G_b[\mathbf{P},\mathbf{\Phi},\mathbf{\sigma}] = U^\sigma_{LGD} + U^\sigma_{Elastic} + U^\sigma_{P\Phi} + U_{A\Phi}, \quad \text{(S1.1a)}$$

$$U^m_{LGD} = a^m_i P_i^2 + a^m_{ij} P_i^2 P_j^2 + \frac{g^m_{ijkl}}{2}\frac{\partial P_i}{\partial x_j}\frac{\partial P_k}{\partial x_l} + b^m_i \Phi_i^2 + b^m_{ij}\Phi_i^2\Phi_j^2 - \eta^m_{ijkl} P_i P_j \Phi_k \Phi_l + \frac{v^m_{ijkl}}{2}\frac{\partial \Phi_i}{\partial x_j}\frac{\partial \Phi_k}{\partial x_l}. \quad \text{(S1.1b)}$$

Superscript $m = u$ or $\sigma$.

$$U^u_{Elastic} = -q_{ijkl} u_{ij} P_k P_l + \frac{f_{ijkl}}{2}\left(\frac{\partial P_k}{\partial x_l}u_{ij} - P_k\frac{\partial u_{ij}}{\partial x_l}\right) + \frac{c_{ijkl}}{2}u_{ij}u_{kl} - r_{ijkl}u_{ij}\Phi_k\Phi_l \quad \text{(S1.1c)}$$

$$U^\sigma_{Elastic} = -Q_{ijkl}\sigma_{ij}P_k P_l + \frac{F_{ijkl}}{2}\left(\frac{\partial P_k}{\partial x_l}\sigma_{ij} - P_k\frac{\partial \sigma_{ij}}{\partial x_l}\right) - \frac{s_{ijkl}}{2}\sigma_{ij}\sigma_{kl} - R_{ijkl}\sigma_{ij}\Phi_k\Phi_l \quad \text{(S1.1d)}$$

Polarization components are $P_i$ ($i$=1, 2, 3). $\Phi_i$ is the components of the structural antiferrodistortive order parameter, e.g. an axial tilt vector corresponding to the octahedral rotation angles [1]; $u_{ij}$ and $\sigma_{ij}$ are the strain and stress tensors correspondingly. The summation is performed over all repeated indices. Coefficients $a_i(T)$ and $b_i(T)$ temperature dependence can be fitted with Curie-Weiss law for ferroelectrics, or with Barrett law for improper ferroelectrics. Gradients coefficients $g_{ij}$ and $v_{ij}$ are regarded positive for commensurate multiferroics. Below we will regard $F_b[\mathbf{P},\mathbf{\Phi},\mathbf{u}]$ as $\Phi$-P-u representation and $G_b[\mathbf{P},\mathbf{\Phi},\mathbf{\sigma}]$ as $\Phi$-P-$\sigma$ representation.

It is well-known that Gibbs and Helmholtz energies are different, but the values calculated from equations of state should be the same in both $\Phi$-P-$\sigma$ and $\Phi$-P-u representations. Starting from the variation of the functional in any of representations, Euler-Lagrange equations for the polarization and tilt as well as equations of state for the elastic stress or strain can be derived. The equations of state give the relation between the stress and strain. After the substitution of the relation into the Euler-Lagrange equations, unambiguous relationship between the coefficients of LGD expansion for mechanically clamped $F_b[\mathbf{P},\mathbf{\Phi},\mathbf{u}]$ and free $G_b[\mathbf{P},\mathbf{\Phi},\mathbf{\sigma}]$ systems can be established. They are listed in the **Table S1.**

**Table S1.** Relations between the $\Phi$-P-$\sigma$ and $\Phi$-P-u LGD expansion coefficients

| LGD-expansion coefficient name | Relationship |
|---|---|
| linear inverse susceptibility (stiffness) | $a^u_i \equiv a^\sigma_i \equiv a_i$, $b^u_i \equiv b^\sigma_i \equiv b_i$ |

| Nonlinear dielectric stiffness | $a^u_{ijkl} = a^\sigma_{ijkl} + Q_{ijmn}q_{mnkl}/2$ |
| --- | --- |
| Nonlinear tilt expansion coefficients | $b^u_{ijkl} = b^\sigma_{ijkl} + R_{ijmn}r_{mnkl}/2$ |
| Polarization gradient coefficient | $g^u_{ijkl} = g^\sigma_{ijkl} + F_{ijmn}f_{mnkl}$ |
| Tilt gradient coefficient | $v^u_{ijkl} \equiv v^\sigma_{ijkl} \equiv v_{ijkl}$ |
| Electrostriction tensor | $q_{ijmn} = Q_{ijkl}c_{klmn}$ |
| Rotostriction tensor | $r_{ijmn} = R_{ijkl}c_{klmn}$ |
| Flexoelectric coupling tensor | $f_{ijmn} = F_{ijkl}c_{klmn}$ |
| Elastic constants | $s_{ijkl}c_{klmn} = (\delta_{im}\delta_{jn} + \delta_{in}\delta_{jm})/2$ |
| Biquadratic coupling between the tilt and polarization | $\eta^u_{ijkl} = \eta^\sigma_{ijkl} - Q_{ijmn}r_{mnkl}$ |
| Flexo-antiferrodistortive coupling between the tilt and polarization gradient | $\xi^u_{ijkl} = \xi^\sigma_{ijkl} + F_{ijmn}r_{mnkl}$ |
| Direct coupling between the polarization components and gradients **not included in Eq.( S1.1)** | $\chi^u_{ijkl} = \chi^\sigma_{ijkl} + F_{ijmn}q_{mnkl}$ |

Allowing for the flexoelectric coupling, one can see the *gap* when compare the relations between the tensorial coefficients in different presentations, summarized in the pre-last row of the **Table S1**. Actually the flexo-antiferrodistortive coupling $\xi^\sigma_{ijkl}$ between the tilt and polarization gradient is allowed by the symmetry. If one starts from conventional Φ-P-σ representation with zero $\xi^\sigma_{ijkl} \equiv 0$, then mandatory come to nonzero values $\xi^u_{ijkl} \equiv F_{ijmn}r_{mnkl}$ in Φ-P-σ representation, i.e. the novel coupling tensor, which strength is proportional to the convolution of the flexoelectric and rotostriction coupling tensors, appear due to the roto-flexo effect. However, it is not enough grounded to set $\xi^\sigma_{ijkl}$ zero, since the components of the direct flexo-distortive coupling tensor are unknown and cannot be determined solely from the LGD-phenomenology, rather they should be either calculated from the first principles or determined experimentally. To summarize, the direct flexo-antiferrodistortive coupling between the polarization gradient and tilt components product should be included in the functionals (S1.1b) in the form of Lifshitz invariant:

$$U_\xi[\mathbf{P},\mathbf{\Phi}] = \frac{\xi^{u,\sigma}_{ijkl}}{2}\left(\Phi_i\Phi_j\frac{\partial P_k}{\partial x_l} - P_k\frac{\partial(\Phi_i\Phi_j)}{\partial x_l}\right) \quad (S1.2)$$

The relationship $\xi^u_{ijkl} = \xi^\sigma_{ijkl} + F_{ijmn}r_{mnkl}$ (see the pre-last raw in the **Table S1**). Similarly the polarization gradient terms $\chi^u_{ijkl}P_iP_j\frac{\partial P_k}{\partial \widetilde{x}_l}$ and $\chi^\sigma_{ijkl}P_iP_j\frac{\partial P_k}{\partial \widetilde{x}_l}$ are allowed by the symmetry. Corresponding relationship is $\chi^u_{ijkl} = \chi^\sigma_{ijkl} + F_{ijmn}q_{mnkl}$ (see the last raw in the **Table S1**). Using **Table S1**, we could write the relations in the evident form

$$a^u_{11} = a^\sigma_{11} + \frac{1}{2}(Q_{11}q_{11} + 2Q_{12}q_{12}), \quad a^u_{12} = a^\sigma_{12} + Q_{11}q_{12} + Q_{12}q_{11} + Q_{12}q_{12} + \frac{1}{2}Q_{44}q_{44} \quad (S1.3a)$$

$$a_{11}^\sigma - a_{11}^u = -\frac{(Q_{11}-Q_{12})^2}{3(s_{11}-s_{12})} - \frac{(Q_{11}+2Q_{12})^2}{6(s_{11}+2s_{12})}, \quad a_{12}^\sigma - a_{12}^u = \frac{(Q_{11}-Q_{12})^2}{3(s_{11}-s_{12})} - \frac{(Q_{11}+2Q_{12})^2}{3(s_{11}+2s_{12})} - \frac{Q_{44}^2}{2s_{44}} \quad \text{(S1.3b)}$$

$$b_{11}^u = b_{11}^\sigma + \frac{1}{2}(R_{11}r_{11} + 2R_{12}r_{12}), \quad b_{12}^u = b_{12}^\sigma + R_{11}r_{12} + R_{12}r_{11} + R_{12}r_{12} + \frac{1}{2}R_{44}r_{44} \quad \text{(S1.4a)}$$

$$b_{11}^\sigma = b_{11}^u - \frac{(r_{11}-r_{12})^2}{3(c_{11}-c_{12})} - \frac{(r_{11}+2r_{12})^2}{6(c_{11}+2c_{12})} \quad b_{12}^\sigma = b_{12}^u + \frac{(r_{11}-r_{12})^2}{3(c_{11}-c_{12})} - \frac{(r_{11}+2r_{12})^2}{3(c_{11}+2c_{12})} - \frac{r_{44}^2}{2c_{44}} \quad \text{(S1.4b)}$$

$$\eta_{11}^u = \eta_{11}^\sigma + (Q_{11}r_{11} + 2Q_{12}r_{12}), \quad \eta_{12}^u = \eta_{12}^\sigma + Q_{11}r_{12} + Q_{12}r_{11} + Q_{12}r_{12}, \quad \eta_{44}^u = \eta_{44}^\sigma + Q_{44}r_{44} \quad \text{(S1.5a)}$$

$$\eta_{11}^\sigma = \eta_{11}^u + \frac{2(q_{11}-q_{12})(r_{11}-r_{12})}{3(c_{11}-c_{12})} + \frac{(q_{11}+2q_{12})(r_{11}+2r_{12})}{3(c_{11}+2c_{12})} \quad \text{(S1.5b)}$$

$$\eta_{12}^\sigma = \eta_{12}^u - \frac{(q_{11}-q_{12})(r_{11}-r_{12})}{3(c_{11}-c_{12})} + \frac{(q_{11}+2q_{12})(r_{11}+2r_{12})}{3(c_{11}+2c_{12})} \quad \eta_{44}^\sigma = \eta_{44}^u + \frac{q_{44}r_{44}}{c_{44}} \quad \text{(S1.5c)}$$

Note the relationships between "strain" and "stress" coefficients:

$$Q_{11} = \frac{(c_{11}+c_{12})q_{11} - 2c_{12}q_{12}}{(c_{11}-c_{12})(c_{11}+2c_{12})}, \quad Q_{12} = \frac{c_{11}q_{12} - c_{12}q_{11}}{(c_{11}-c_{12})(c_{11}+2c_{12})}, \quad Q_{44} = \frac{q_{44}}{c_{44}}, \quad \text{(S1.6)}$$

$$s_{11} = \frac{c_{11}+c_{12}}{(c_{11}-c_{12})(c_{11}+2c_{12})}, \quad s_{12} = \frac{-c_{12}}{(c_{11}-c_{12})(c_{11}+2c_{12})}, \quad s_{44} = \frac{1}{c_{44}}, \quad \text{(S1.7)}$$

In some cases the following relations between tensor invariants could be useful:

$$c_{11} - c_{12} = \frac{1}{s_{11}-s_{12}}, \quad c_{11} + 2c_{12} = \frac{1}{s_{11}+2s_{12}} \quad \text{(S1.8)}$$

$$q_{11} - q_{12} = (Q_{11}-Q_{12})(c_{11}-c_{12}) = \frac{Q_{11}-Q_{12}}{s_{11}-s_{12}}, \quad q_{11} + 2q_{12} = (Q_{11}+2Q_{12})(c_{11}+2c_{12}) = \frac{Q_{11}+2Q_{12}}{s_{11}+2s_{12}}. \quad \text{(S1.9a)}$$

$$\frac{(q_{11}-q_{12})^2}{c_{11}-c_{12}} = \frac{(Q_{11}-Q_{12})^2}{s_{11}-s_{12}}, \quad \frac{(q_{11}+2q_{12})^2}{c_{11}+2c_{12}} = \frac{(Q_{11}+2Q_{12})^2}{s_{11}+2s_{12}} \quad \text{(S1.9b)}$$

### S1.2. Transformation laws of Eq.(2b)

**a)** Transformation laws of **pseudo**-tensor $\zeta_{ijk}$, true vector **A**, pseudo-vector $\Phi$ and coordinate derivative $\partial/\partial x_l$ are:

$$\widetilde{\zeta}_{ijk} = \det(B) B_{im} B_{jg} B_{ks} \zeta_{mgs}, \quad \text{(S1.10a)}$$

$$\widetilde{A}_i = B_{ip} A_p, \quad \widetilde{\Phi}_k = \det(B) B_{kf} \Phi_f, \quad \frac{\partial}{\partial \widetilde{x}_j} = B_{jl} \frac{\partial}{\partial x_l}. \quad \text{(S1.10b)}$$

Here the summation is performed over the repeating indexes. **B** is the unitary transformation matrix with components $B_{ij}$ ($i,j = 1,2,3$) with convolution $B_{ij} B_{ik} = \delta_{jk}$ and determinant $\det(\mathbf{B}) = \pm 1$. For the case when the matrices **B** represent all the elements of the parent phase point symmetry group

Eqs.(A.1) become identity without symbol "tilda", e.g. one could obtain the system of linear equations $\zeta_{ijk} = \det(B) B_{im} B_{jg} B_{ks} \zeta_{mgs}$. The case will be considered hereinafter

**b)** Translations rules of **A** and **Φ** corresponding permutation ($a \leftrightarrow b$) of the equivalent cells $a$ and $b$ are

$$\mathbf{A} \leftrightarrow -\mathbf{A} \text{ simultaneously with } \mathbf{\Phi} \leftrightarrow -\mathbf{\Phi} \tag{S1.11}$$

The first rule $\mathbf{A} \leftrightarrow -\mathbf{A}$ follows from the definition of $\mathbf{A} = (\mathbf{P}_a - \mathbf{P}_b)/2$, since $\mathbf{P}_a \leftrightarrow \mathbf{P}_b$ with $a \leftrightarrow b$. The second rule $\mathbf{\Phi} \leftrightarrow -\mathbf{\Phi}$ follows from the definition of $\mathbf{\Phi} = (\mathbf{\Phi}_a - \mathbf{\Phi}_b)/2$.

Elementary calculation listed below proves that the invariant (1) (*main text*) is indeed invariant with respect to the transformation of parent phase point symmetry group using the elements **B** of the and with respect to the permutation operation $a \leftrightarrow b$.

a) Invariance of expression (1) with respect to the transformation **B** can be proved directly:

$$\widetilde{\zeta}_{ijk} \widetilde{A}_i \frac{\partial \widetilde{\Phi}_k}{\partial \widetilde{x}_j} = \det^2(B) B_{im} B_{jg} B_{ks} \zeta_{mgs} B_{ip} A_p B_{jl} \frac{\partial}{\partial x_l} B_{kf} \Phi_f =$$
$$= B_{im} B_{ip} B_{jg} B_{jl} B_{ks} B_{kf} \zeta_{mgs} A_p \frac{\partial \Phi_f}{\partial x_l} = \delta_{mp} \delta_{gl} \delta_{kf} \zeta_{mgs} A_p \frac{\partial \Phi_f}{\partial x_l} = e_{plf} A_p \frac{\partial \Phi_f}{\partial x_l} \tag{S1.12}$$

Derivation (S1.12) becomes trivial for cubic symmetry. Because **Φ** is an axial vector [2], rot**Φ** becomes "true" polar vector. Similarly because **A** is polar vector, rot**A** is an axial vector. The scalar product of two vectors $\mathbf{A} \cdot \text{rot} \mathbf{\Phi}$ (as well as two pseudovectors $\mathbf{\Phi} \cdot \text{rot} \mathbf{A}$) becomes a true scalar.

b) Invariance with respect to the permutation operation is evident once Eq.(2) is written in the full form, $U_{A\Phi}[\mathbf{A}, \mathbf{\Phi}] = \frac{\zeta_{ijk}}{2} \left( (P_{ai} - P_{bi}) \frac{\partial (\Phi_{ak} - \Phi_{bk})}{\partial x_j} - (\Phi_{ak} - \Phi_{bk}) \frac{\partial (P_{ai} - P_{bi})}{\partial x_j} \right)$. For cubic parent symmetry $U_{A\Phi} = \frac{\zeta}{8} ((\mathbf{P}_a - \mathbf{P}_b) \cdot \text{rot}(\mathbf{\Phi}_a - \mathbf{\Phi}_b) - (\mathbf{\Phi}_a - \mathbf{\Phi}_b) \cdot \text{rot}(\mathbf{P}_a - \mathbf{P}_b))$.

Since the invariant (2b) is nonzero in the antiferroelectric-antiferrodistortive phase, one may ask a reasonable question about similar invariant relevant for the appearance of incommensurate ferroelectric-antiferrodistortive phase with nonzero polarization $\mathbf{P} = (\mathbf{P}_a + \mathbf{P}_b)/2$. Note that the permutative-symmetric term $U_P = \frac{\zeta^*_{ijk}}{8} \left( (P_{ai} + P_{bi}) \frac{\partial (\Phi_{ak} + \Phi_{bk})}{\partial x_j} - (\Phi_{ak} + \Phi_{bk}) \frac{\partial (P_{ai} + P_{bi})}{\partial x_j} \right)$ is identically zero in the antiferrodistortive phase with $\mathbf{\Phi}_a = -\mathbf{\Phi}_b$. The coupling terms like $U_P \propto \frac{\zeta^*_{ijk}}{8} \left( (P_{ai} + P_{bi}) \frac{\partial (\Phi_{ak} - \Phi_{bk})}{\partial x_j} - (\Phi_{ak} - \Phi_{bk}) \frac{\partial (P_{ai} + P_{bi})}{\partial x_j} \right)$ are permutative-antisymmetric and thus forbidden.

# Appendix S2

## Equations for 90-degree TB in atiferrodistortive phases in Φ-P-σ representation

For the considered case of 1D distribution near [110] TB the free energy density in the rotated frame is

$$\begin{aligned}
G = {} & a_1(T)\left(\widetilde{P}_1^2 + \widetilde{P}_2^2\right) + \widetilde{a}_{11}^\sigma\left(\widetilde{P}_1^4 + \widetilde{P}_2^4\right) + \widetilde{a}_{12}^\sigma\left(\widetilde{P}_1^2 \widetilde{P}_2^2\right) - \widetilde{P}_i \widetilde{E}_i^{ext} - \frac{1}{2}\widetilde{P}_1 \widetilde{E}_1^d \\
& - \widetilde{Q}_{11}\left(\widetilde{\sigma}_{11}\widetilde{P}_1^2 + \widetilde{\sigma}_{22}\widetilde{P}_2^2\right) - \widetilde{Q}_{12}\left(\widetilde{\sigma}_{11}\widetilde{P}_2^2 + \widetilde{\sigma}_{22}\widetilde{P}_1^2\right) - Q_{12}\widetilde{\sigma}_{33}\left(\widetilde{P}_1^2 + \widetilde{P}_2^2\right) - \widetilde{Q}_{66}\widetilde{\sigma}_{12}\widetilde{P}_1\widetilde{P}_2 \\
& - \frac{1}{2}\widetilde{s}_{11}\left(\widetilde{\sigma}_{11}^2 + \widetilde{\sigma}_{22}^2\right) - \frac{1}{2}s_{11}\widetilde{\sigma}_{33}^2 - \widetilde{s}_{12}\widetilde{\sigma}_{11}\widetilde{\sigma}_{22} - s_{12}\left(\widetilde{\sigma}_{22}\widetilde{\sigma}_{33} + \widetilde{\sigma}_{33}\widetilde{\sigma}_{11}\right) - \frac{1}{2}\widetilde{s}_{66}\widetilde{\sigma}_{12}^2 - \frac{1}{2}s_{44}\left(\widetilde{\sigma}_{23}^2 + \widetilde{\sigma}_{13}^2\right) \\
& + \beta_1(T)\left(\widetilde{\Phi}_1^2 + \widetilde{\Phi}_2^2\right) + \widetilde{\beta}_{11}^\sigma\left(\widetilde{\Phi}_1^4 + \widetilde{\Phi}_2^4\right) + \widetilde{\beta}_{12}^\sigma\widetilde{\Phi}_1^2\widetilde{\Phi}_2^2 \\
& - \widetilde{R}_{11}\left(\widetilde{\sigma}_{11}\widetilde{\Phi}_1^2 + \widetilde{\sigma}_{22}\widetilde{\Phi}_2^2\right) - \widetilde{R}_{12}\left(\widetilde{\sigma}_{11}\widetilde{\Phi}_2^2 + \widetilde{\sigma}_{22}\widetilde{\Phi}_1^2\right) - R_{12}\widetilde{\sigma}_{33}\left(\widetilde{\Phi}_1^2 + \widetilde{\Phi}_2^2\right) - \widetilde{R}_{66}\left(\widetilde{\sigma}_{12}\widetilde{\Phi}_1\widetilde{\Phi}_2\right) - \\
& - \widetilde{\eta}_{11}^\sigma\left(\widetilde{\Phi}_1^2\widetilde{P}_1^2 + \widetilde{\Phi}_2^2\widetilde{P}_2^2\right) - \widetilde{\eta}_{12}^\sigma\left(\widetilde{\Phi}_1^2\widetilde{P}_2^2 + \widetilde{\Phi}_2^2\widetilde{P}_1^2\right) - \widetilde{\eta}_{66}^\sigma\left(\widetilde{\Phi}_1\widetilde{\Phi}_2\widetilde{P}_1\widetilde{P}_2\right) \\
& + \frac{\widetilde{v}_{11}}{2}\left(\frac{\partial\widetilde{\Phi}_1}{\partial\widetilde{x}_1}\right)^2 + \frac{\widetilde{v}_{66}}{2}\left(\frac{\partial\widetilde{\Phi}_2}{\partial\widetilde{x}_1}\right)^2 + \frac{\widetilde{g}_{11}^\sigma}{2}\left(\frac{\partial\widetilde{P}_1}{\partial\widetilde{x}_1}\right)^2 + \frac{\widetilde{g}_{66}^\sigma}{2}\left(\frac{\partial\widetilde{P}_2}{\partial\widetilde{x}_1}\right)^2 \\
& + \left(\widetilde{F}_{11}\widetilde{\sigma}_{11} + \widetilde{F}_{12}\widetilde{\sigma}_{22} + F_{12}\widetilde{\sigma}_{33}\right)\frac{\partial\widetilde{P}_1}{\partial\widetilde{x}_1} + \widetilde{F}_{66}\widetilde{\sigma}_{12}\frac{\partial\widetilde{P}_2}{\partial\widetilde{x}_1}
\end{aligned} \quad \text{(S2.1)}$$

Here $\widetilde{E}_i^{ext}$ and $\widetilde{E}_i^d$ are the components of external and depolarization field respectively.

Using compatibility and mechanical equilibrium conditions the case of $\widetilde{x}_1$-dependent solution we obtained the evident form of elastic strains and stresses in rotated system:

$$\widetilde{\sigma}_{22} = \frac{s_{11}U_2 - s_{12}U_3}{s_{11}\widetilde{s}_{11} - s_{12}^2}, \qquad \widetilde{\sigma}_{33} = \frac{\widetilde{s}_{11}U_3 - s_{12}U_2}{s_{11}\widetilde{s}_{11} - s_{12}^2}, \qquad \widetilde{\sigma}_{11} = \widetilde{\sigma}_{13} = \widetilde{\sigma}_{12} = \widetilde{\sigma}_{23} = 0, \quad \text{(S2.2a)}$$

$$\widetilde{u}_{22} = \left(\widetilde{R}_{11} + \widetilde{R}_{12}\right)\frac{\Phi_S^2}{2}, \qquad \widetilde{u}_{33} = R_{12}\Phi_S^2, \quad \text{(S2.2b)}$$

$$\widetilde{u}_{11} = -\widetilde{F}_{11}\frac{\partial\widetilde{P}_1}{\partial\widetilde{x}_1} + \widetilde{Q}_{11}\widetilde{P}_1^2 + \widetilde{Q}_{12}\widetilde{P}_2^2 + \widetilde{R}_{11}\widetilde{\Phi}_1^2 + \widetilde{R}_{12}\widetilde{\Phi}_2^2 + \frac{\left(\widetilde{s}_{12}s_{11} - s_{12}^2\right)U_2 + \left(s_{12}\widetilde{s}_{11} - \widetilde{s}_{12}s_{12}\right)U_3}{s_{11}\widetilde{s}_{11} - s_{12}^2}, \quad \text{(S2.2c)}$$

$$\widetilde{u}_{12} = \frac{1}{2}\left(-\widetilde{F}_{66}\frac{\partial\widetilde{P}_2}{\partial\widetilde{x}_1} + \widetilde{Q}_{66}\widetilde{P}_1\widetilde{P}_2 + \widetilde{R}_{66}\widetilde{\Phi}_1\widetilde{\Phi}_2\right), \quad \widetilde{u}_{13} \equiv 0, \ \widetilde{u}_{23} \equiv 0. \quad \text{(S2.2d)}$$

Functions:

$$U_2 = \widetilde{F}_{12}\frac{\partial\widetilde{P}_1}{\partial\widetilde{x}_1} + \widetilde{R}_{11}\left(\frac{\Phi_S^2}{2} - \widetilde{\Phi}_2^2\right) + \widetilde{R}_{12}\left(\frac{\Phi_S^2}{2} - \widetilde{\Phi}_1^2\right) - \widetilde{Q}_{12}\widetilde{P}_1^2 - \widetilde{Q}_{11}\widetilde{P}_2^2 \quad \text{(S2.2e)}$$

$$U_3 = F_{12}\frac{\partial\widetilde{P}_1}{\partial\widetilde{x}_1} + R_{12}\left(\Phi_S^2 - \widetilde{\Phi}_2^2 - \widetilde{\Phi}_1^2\right) - Q_{12}\left(\widetilde{P}_2^2 + \widetilde{P}_1^2\right) \quad \text{(S2.2f)}$$

In these equations we used the tensor components in the new reference frame for elastic compliances $\widetilde{s}_{11} = (s_{11} + s_{12} + s_{44}/2)/2$, $\widetilde{s}_{12} = (s_{11} + s_{12} - s_{44}/2)/2$, $\widetilde{s}_{66} = 2(s_{11} - s_{12})$; rotostriction

$\widetilde{R}_{11} = (R_{11} + R_{12} + R_{44}/2)/2$ , $\widetilde{R}_{12} = (R_{11} + R_{12} - R_{44}/2)/2$ , $\widetilde{R}_{66} = 2(R_{11} - R_{12})$ ; electrostriction $\widetilde{Q}_{11} = (Q_{11} + Q_{12} + Q_{44}/2)/2$ , $\widetilde{Q}_{12} = (Q_{11} + Q_{12} - Q_{44}/2)/2$ , $\widetilde{Q}_{66} = 2(Q_{11} - Q_{12})$ ; flexoelectric coefficients $\widetilde{F}_{11} = (F_{11} + F_{12} + F_{44})/2$ , $\widetilde{F}_{12} = (F_{11} + F_{12} - F_{44})/2$ , $\widetilde{F}_{66} = F_{11} - F_{12}$. Since polarization components are tends to zero far from the twin wall, the stresses (S2.2a) vanish far from the wall; the strains (S2.2b-d) tend to the spontaneous strains.

$$2\left(b_1 - \widetilde{R}_{12}\widetilde{\sigma}_{22} - R_{12}\widetilde{\sigma}_{33} + \widetilde{b}_{12}^\sigma \widetilde{\Phi}_2^2 - \widetilde{\eta}_{11}^\sigma \widetilde{P}_1^2 - \widetilde{\eta}_{12}^\sigma \widetilde{P}_2^2\right)\widetilde{\Phi}_1 + 4\widetilde{b}_{11}^\sigma \widetilde{\Phi}_1^3 - \widetilde{\eta}_{66}^\sigma \widetilde{P}_1 \widetilde{P}_2 \widetilde{\Phi}_2 - \widetilde{v}_{11}\frac{\partial^2 \widetilde{\Phi}_1}{\partial \widetilde{x}_1^2} = 0, \quad \text{(S2.3a)}$$

$$2\left(b_1 - \widetilde{R}_{11}\widetilde{\sigma}_{22} - R_{12}\widetilde{\sigma}_{33} + \widetilde{b}_{12}^\sigma \widetilde{\Phi}_1^2 - \widetilde{\eta}_{11}^\sigma \widetilde{P}_2^2 - \widetilde{\eta}_{12}^\sigma \widetilde{P}_1^2\right)\widetilde{\Phi}_2 + 4\widetilde{b}_{11}^\sigma \widetilde{\Phi}_2^3 - \widetilde{\eta}_{66}^\sigma \widetilde{P}_1 \widetilde{P}_2 \widetilde{\Phi}_1 - \widetilde{v}_{66}\frac{\partial^2 \widetilde{\Phi}_2}{\partial \widetilde{x}_1^2} = 0, \quad \text{(S2.3b)}$$

$$2\left(a_1 - \widetilde{Q}_{12}\widetilde{\sigma}_{22} - Q_{12}\widetilde{\sigma}_{33} + \widetilde{a}_{12}^\sigma \widetilde{P}_2^2 - \widetilde{\eta}_{11}^\sigma \widetilde{\Phi}_1^2 - \widetilde{\eta}_{12}^\sigma \widetilde{\Phi}_2^2\right)\widetilde{P}_1 + 4\widetilde{a}_{11}^\sigma \widetilde{P}_1^3 - \widetilde{\eta}_{66}^\sigma \widetilde{\Phi}_2 \widetilde{\Phi}_1 \widetilde{P}_2 - \widetilde{g}_{11}^\sigma \frac{\partial^2 \widetilde{P}_1}{\partial \widetilde{x}_1^2} =$$
$$= \widetilde{E}_1^{ext} + \widetilde{E}_1^d + \widetilde{F}_{12}\frac{\partial \widetilde{\sigma}_{22}}{\partial \widetilde{x}_1} + F_{12}\frac{\partial \widetilde{\sigma}_{33}}{\partial \widetilde{x}_1} \quad \text{(S2.3c)}$$

$$2\left(a_1 - \widetilde{Q}_{11}\widetilde{\sigma}_{22} - Q_{12}\widetilde{\sigma}_{33} + \widetilde{a}_{12}^\sigma \widetilde{P}_1^2 - \widetilde{\eta}_{11}^\sigma \widetilde{\Phi}_2^2 - \widetilde{\eta}_{12}^\sigma \widetilde{\Phi}_1^2\right)\widetilde{P}_2 + 4\widetilde{a}_{11}^\sigma \widetilde{P}_2^3 - \widetilde{\eta}_{66}^\sigma \widetilde{\Phi}_1 \widetilde{\Phi}_2 \widetilde{P}_1 - \widetilde{g}_{66}^\sigma \frac{\partial^2 \widetilde{P}_2}{\partial \widetilde{x}_1^2} = E_2^{ext}.$$

(S2.3d)

Here we introduced the tensor components in the new reference frame:

$\widetilde{b}_{11}^\sigma = \frac{1}{4}(2b_{11}^\sigma + b_{12}^\sigma)$ , $\widetilde{b}_{12}^\sigma = \frac{1}{2}(6b_{11}^\sigma - b_{12}^\sigma)$ , $\widetilde{a}_{11}^\sigma = \frac{1}{4}(2a_{11}^\sigma + a_{12}^\sigma)$ , $\widetilde{a}_{12}^\sigma = \frac{1}{2}(6a_{11}^\sigma - a_{12}^\sigma)$ ,

$\widetilde{\eta}_{11}^\sigma = (\eta_{11}^\sigma + \eta_{12}^\sigma + \eta_{44}^\sigma/2)/2$ , $\widetilde{\eta}_{12}^\sigma = (\eta_{11}^\sigma + \eta_{12}^\sigma - \eta_{44}^\sigma/2)/2$ , $\widetilde{\eta}_{66}^\sigma = 2(\eta_{11}^\sigma - \eta_{12}^\sigma)$ ,

$\widetilde{v}_{11} = (v_{11} + v_{12} + 2v_{44})/2$, $\widetilde{v}_{66} = (v_{11} - v_{12})/2$, $\widetilde{g}_{11}^\sigma = (g_{11}^\sigma + g_{12}^\sigma + 2g_{44}^\sigma)/2$, $\widetilde{g}_{66}^\sigma = (g_{11}^\sigma - g_{12}^\sigma)/2$.

Substitution of the elastic solution (S2.2) in Eqs.(S2.3) leads to the system of four coupled equations for the tilt and polarization vectors components:

$$2\left(b_1 - \widetilde{R}_{12}\frac{s_{11}U_2 - s_{12}U_3}{s_{11}\widetilde{s}_{11} - s_{12}^2} - R_{12}\frac{\widetilde{s}_{11}U_3 - s_{12}U_2}{s_{11}\widetilde{s}_{11} - s_{12}^2} + \widetilde{b}_{12}^\sigma \widetilde{\Phi}_2^2 - \widetilde{\eta}_{11}^\sigma \widetilde{P}_1^2 - \widetilde{\eta}_{12}^\sigma \widetilde{P}_2^2\right)\widetilde{\Phi}_1$$
$$+ 4\widetilde{b}_{11}^\sigma \widetilde{\Phi}_1^3 - \widetilde{\eta}_{66}^\sigma \widetilde{P}_1 \widetilde{P}_2 \widetilde{\Phi}_2 - \widetilde{v}_{11}\frac{\partial^2 \widetilde{\Phi}_1}{\partial \widetilde{x}_1^2} = 0 \quad \text{(S2.4a)}$$

$$2\left(b_1 - \widetilde{R}_{11}\frac{s_{11}U_2 - s_{12}U_3}{s_{11}\widetilde{s}_{11} - s_{12}^2} - R_{12}\frac{\widetilde{s}_{11}U_3 - s_{12}U_2}{s_{11}\widetilde{s}_{11} - s_{12}^2} + \widetilde{b}_{12}^\sigma \widetilde{\Phi}_1^2 - \widetilde{\eta}_{11}^\sigma \widetilde{P}_2^2 - \widetilde{\eta}_{12}^\sigma \widetilde{P}_1^2\right)\widetilde{\Phi}_2$$
$$+ 4\widetilde{b}_{11}^\sigma \widetilde{\Phi}_2^3 - \widetilde{\eta}_{66}^\sigma \widetilde{P}_1 \widetilde{P}_2 \widetilde{\Phi}_1 - \widetilde{v}_{66}\frac{\partial^2 \widetilde{\Phi}_2}{\partial \widetilde{x}_1^2} = 0 \quad \text{(S2.4b)}$$

$$2\left(a_1 - \tilde{Q}_{12}\frac{s_{11}U_2 - s_{12}U_3}{s_{11}\tilde{s}_{11} - s_{12}^2} - Q_{12}\frac{\tilde{s}_{11}U_3 - s_{12}U_2}{s_{11}\tilde{s}_{11} - s_{12}^2} + \tilde{a}_{12}^\sigma \tilde{P}_2^2 - \tilde{\eta}_{11}^\sigma \tilde{\Phi}_1^2 - \tilde{\eta}_{12}^\sigma \tilde{\Phi}_2^2\right)\tilde{P}_1 + 4\tilde{a}_{11}^\sigma \tilde{P}_1^3$$
$$-\tilde{\eta}_{66}^\sigma \tilde{\Phi}_2 \tilde{\Phi}_1 \tilde{P}_2 - \tilde{g}_{11}^\sigma \frac{\partial^2 \tilde{P}_1}{\partial \tilde{x}_1^2} = \tilde{E}_1^{ext} + \tilde{E}_1^d + \tilde{F}_{12}\frac{\partial}{\partial \tilde{x}_1}\left(\frac{s_{11}U_2 - s_{12}U_3}{s_{11}\tilde{s}_{11} - s_{12}^2}\right) + F_{12}\frac{\partial}{\partial \tilde{x}_1}\left(\frac{\tilde{s}_{11}U_3 - s_{12}U_2}{s_{11}\tilde{s}_{11} - s_{12}^2}\right),$$ (S2.4c)

$$2\left(a_1 + \tilde{a}_{12}^\sigma \tilde{P}_1^2 - \tilde{\eta}_{11}^\sigma \tilde{\Phi}_2^2 - \tilde{\eta}_{12}^\sigma \tilde{\Phi}_1^2 - \tilde{Q}_{11}\frac{s_{11}U_2 - s_{12}U_3}{s_{11}\tilde{s}_{11} - s_{12}^2} - Q_{12}\frac{\tilde{s}_{11}U_3 - s_{12}U_2}{s_{11}\tilde{s}_{11} - s_{12}^2}\right)\tilde{P}_2$$
$$+ 4\tilde{a}_{11}^\sigma \tilde{P}_2^3 - \tilde{\eta}_{66}^\sigma \tilde{\Phi}_1 \tilde{\Phi}_2 \tilde{P}_1 - \tilde{g}_{66}^\sigma \frac{\partial^2 \tilde{P}_2}{\partial \tilde{x}_1^2} = E_2^{ext}$$ (S2.4d)

The terms $\tilde{F}_{12}\frac{\partial}{\partial \tilde{x}_1}\left(\frac{s_{11}U_2 - s_{12}U_3}{s_{11}\tilde{s}_{11} - s_{12}^2}\right) + F_{12}\frac{\partial}{\partial \tilde{x}_1}\left(\frac{\tilde{s}_{11}U_3 - s_{12}U_2}{s_{11}\tilde{s}_{11} - s_{12}^2}\right)$ in the right-hand-side of Eq.(S2.4c) consist of the terms proportional to gradient of polarization powers and flexo-roto field:

$$\tilde{E}_1^{FR} = \frac{\tilde{F}_{12}(s_{12}R_{12} - s_{11}\tilde{R}_{11}) + F_{12}(s_{12}\tilde{R}_{11} - \tilde{s}_{11}R_{12})}{s_{11}\tilde{s}_{11} - s_{12}^2}\frac{\partial(\tilde{\Phi}_2^2)}{\partial \tilde{x}_1} +$$
$$\frac{\tilde{F}_{12}(s_{12}R_{12} - s_{11}\tilde{R}_{12}) + F_{12}(s_{12}\tilde{R}_{12} - \tilde{s}_{11}R_{12})}{s_{11}\tilde{s}_{11} - s_{12}^2}\frac{\partial(\tilde{\Phi}_1^2)}{\partial \tilde{x}_1}$$ (S2.5)

$\tilde{E}_1^{FR}(\tilde{x}_1)$ is an odd function of coordinate.

Boundary conditions for the tilt vector at **hard** twins (with rotation vector **parallel** to the wall plane in the immediate of the wall, **Fig.1a**) are

$$\tilde{\Phi}_1(\tilde{x}_1 = 0) = 0, \quad \frac{\partial \tilde{\Phi}_2}{\partial \tilde{x}_1}(\tilde{x}_1 = 0) = 0$$ (S2.6)

at the TB. Very far from TB:

$$\tilde{\Phi}_1(\tilde{x}_1 \to +\infty) = \frac{\Phi_S}{\sqrt{2}}, \quad \tilde{\Phi}_1(\tilde{x}_1 \to -\infty) = -\frac{\Phi_S}{\sqrt{2}}, \quad \tilde{\Phi}_2(\tilde{x}_1 \to \pm\infty) = \frac{\Phi_S}{\sqrt{2}},$$ (S2.6b)

Boundary conditions for **easy** twins (with rotation vector **perpendicular** to the wall plane in the immediate of the wall, **Fig.1b**) are

$$\frac{\partial \tilde{\Phi}_1}{\partial \tilde{x}_1}(\tilde{x}_1 = 0) = 0, \quad \tilde{\Phi}_2(\tilde{x}_1 = 0) = 0$$ (S2.7a)

at the TB. Very far from for TB:

$$\tilde{\Phi}_1(\tilde{x}_1 \to \pm\infty) = \frac{\Phi_S}{\sqrt{2}}, \quad \tilde{\Phi}_2(\tilde{x}_1 \to +\infty) = \frac{\Phi_S}{\sqrt{2}}, \quad \tilde{\Phi}_2(\tilde{x}_1 \to -\infty) = -\frac{\Phi_S}{\sqrt{2}}$$ (S2.7b)

After the substitution of the elastic solution (S2.2) into the free energy and corresponding Legendre transformation the evident form of thermodynamic potential could be found as

$$F = A_1\widetilde{P}_1^2 + A_2\widetilde{P}_2^2 + A_{11}\widetilde{P}_1^4 + A_{22}\widetilde{P}_2^4 + A_{12}\widetilde{P}_1^2\widetilde{P}_2^2 - \frac{\widetilde{E}_1^d}{2}\widetilde{P}_1 +$$
$$+ B_1\widetilde{\Phi}_1^2 + B_2\widetilde{\Phi}_2^2 + B_{11}\widetilde{\Phi}_1^4 + B_{22}\widetilde{\Phi}_2^4 + B_{12}\widetilde{\Phi}_1^2\widetilde{\Phi}_2^2 -$$
$$- E_{11}\widetilde{\Phi}_1^2\widetilde{P}_1^2 - E_{22}\widetilde{\Phi}_2^2\widetilde{P}_2^2 - E_{12}\widetilde{\Phi}_1^2\widetilde{P}_2^2 - E_{21}\widetilde{\Phi}_2^2\widetilde{P}_1^2 - E_{66}\widetilde{\Phi}_1\widetilde{\Phi}_2\widetilde{P}_1\widetilde{P}_2 \quad \text{(S2.8)}$$
$$+ \frac{\widetilde{v}_{11}}{2}\left(\frac{\partial\widetilde{\Phi}_1}{\partial\widetilde{x}_1}\right)^2 + \frac{\widetilde{v}_{66}}{2}\left(\frac{\partial\widetilde{\Phi}_2}{\partial\widetilde{x}_1}\right)^2 + \frac{G_{11}}{2}\left(\frac{\partial\widetilde{P}_1}{\partial\widetilde{x}_1}\right)^2 + \frac{G_{66}}{2}\left(\frac{\partial\widetilde{P}_2}{\partial\widetilde{x}_1}\right)^2$$
$$+ \left(X_{11}\widetilde{\Phi}_1^2 + X_{21}\widetilde{\Phi}_2^2 + Z_{21}\widetilde{P}_2^2\right)\frac{\partial\widetilde{P}_1}{\partial\widetilde{x}_1}$$

With introduced renormalized coefficients

$$A_1 = a_1 - \left(\frac{R_{12}Q_{12}}{s_{11}} + \frac{\left((s_{11}\widetilde{R}_{11} - s_{12}R_{12}) + (s_{11}\widetilde{R}_{12} - s_{12}R_{12})\right)(s_{11}\widetilde{Q}_{12} - s_{12}Q_{12})}{2s_{11}(s_{11}\widetilde{s}_{11} - s_{12}^2)}\right)\Phi_S^2 \quad \text{(S2.9a)}$$

$$A_2 = a_1 - \left(\frac{R_{12}Q_{12}}{s_{11}} + \frac{\left((s_{11}\widetilde{R}_{11} - s_{12}R_{12}) + (s_{11}\widetilde{R}_{12} - s_{12}R_{12})\right)(s_{11}\widetilde{Q}_{11} - s_{12}Q_{12})}{2s_{11}(s_{11}\widetilde{s}_{11} - s_{12}^2)}\right)\Phi_S^2 \quad \text{(S2.9b)}$$

$$B_1 = b_1 - \left(\frac{R_{12}^2}{s_{11}} + \frac{\left((s_{11}\widetilde{R}_{11} - s_{12}R_{12}) + (s_{11}\widetilde{R}_{12} - s_{12}R_{12})\right)(s_{11}\widetilde{R}_{12} - s_{12}R_{12})}{2s_{11}(s_{11}\widetilde{s}_{11} - s_{12}^2)}\right)\Phi_S^2 \quad \text{(S2.9c)}$$

$$B_2 = b_1 - \left(\frac{R_{12}^2}{s_{11}} + \frac{\left((s_{11}\widetilde{R}_{11} - s_{12}R_{12}) + (s_{11}\widetilde{R}_{12} - s_{12}R_{12})\right)(s_{11}\widetilde{R}_{11} - s_{12}R_{12})}{2s_{11}(s_{11}\widetilde{s}_{11} - s_{12}^2)}\right)\Phi_S^2 \quad \text{(S2.9d)}$$

$$A_{11} = \widetilde{a}_{11}^\sigma + \frac{Q_{12}^2}{2s_{11}} + \frac{(s_{11}\widetilde{Q}_{12} - s_{12}Q_{12})^2}{2s_{11}(s_{11}\widetilde{s}_{11} - s_{12}^2)}, \quad A_{22} = \widetilde{a}_{11}^\sigma + \frac{Q_{12}^2}{2s_{11}} + \frac{(s_{11}\widetilde{Q}_{11} - s_{12}Q_{12})^2}{2s_{11}(s_{11}\widetilde{s}_{11} - s_{12}^2)} \quad \text{(S2.10a)}$$

$$B_{11} = \widetilde{b}_{11}^\sigma + \frac{R_{12}^2}{2s_{11}} + \frac{(s_{11}\widetilde{R}_{12} - s_{12}R_{12})^2}{2s_{11}(s_{11}\widetilde{s}_{11} - s_{12}^2)}, \quad B_{22} = \widetilde{b}_{11}^\sigma + \frac{R_{12}^2}{2s_{11}} + \frac{(s_{11}\widetilde{R}_{11} - s_{12}R_{12})^2}{2s_{11}(s_{11}\widetilde{s}_{11} - s_{12}^2)} \quad \text{(S2.10b)}$$

$$A_{12} = \widetilde{a}_{12}^\sigma + \frac{Q_{12}^2}{s_{11}} + \frac{(s_{11}\widetilde{Q}_{12} - s_{12}Q_{12})(s_{11}\widetilde{Q}_{11} - s_{12}Q_{12})}{s_{11}(s_{11}\widetilde{s}_{11} - s_{12}^2)}, \quad \text{(S2.10c)}$$

$$B_{12} = \widetilde{b}_{12}^\sigma + \frac{R_{12}^2}{s_{11}} + \frac{(s_{11}\widetilde{R}_{12} - s_{12}R_{12})(s_{11}\widetilde{R}_{11} - s_{12}R_{12})}{s_{11}(s_{11}\widetilde{s}_{11} - s_{12}^2)} \quad \text{(S2.10d)}$$

$$E_{11} = \widetilde{\eta}_{11}^\sigma - \frac{R_{12}Q_{12}}{s_{11}} - \frac{(s_{11}\widetilde{Q}_{12} - s_{12}Q_{12})(s_{11}\widetilde{R}_{12} - s_{12}R_{12})}{s_{11}(s_{11}\widetilde{s}_{11} - s_{12}^2)} \quad \text{(S2.11a)}$$

$$E_{12} = \widetilde{\eta}_{12}^\sigma - \frac{R_{12}Q_{12}}{s_{11}} - \frac{(s_{11}\widetilde{Q}_{11} - s_{12}Q_{12})(s_{11}\widetilde{R}_{12} - s_{12}R_{12})}{s_{11}(s_{11}\widetilde{s}_{11} - s_{12}^2)} \quad \text{(S2.11b)}$$

$$E_{22} = \widetilde{\eta}_{11}^\sigma - \frac{R_{12}Q_{12}}{s_{11}} - \frac{(s_{11}\widetilde{Q}_{11} - s_{12}Q_{12})(s_{11}\widetilde{R}_{11} - s_{12}R_{12})}{s_{11}(s_{11}\widetilde{s}_{11} - s_{12}^2)} \quad \text{(S2.11c)}$$

$$E_{21} = \widetilde{\eta}_{12}^\sigma - \frac{R_{12}Q_{12}}{s_{11}} - \frac{(s_{11}\widetilde{Q}_{12} - s_{12}Q_{12})(s_{11}\widetilde{R}_{11} - s_{12}R_{12})}{s_{11}(s_{11}\widetilde{s}_{11} - s_{12}^2)}, \quad E_{66} = \widetilde{\eta}_{66}^\sigma \quad \text{(S2.11d)}$$

Gradient coupling constants are

$$X_{11} = -\frac{F_{12}R_{12}}{s_{11}} - \frac{(s_{11}\widetilde{F}_{12} - s_{12}F_{12})(s_{11}\widetilde{R}_{12} - s_{12}R_{12})}{s_{11}(s_{11}\widetilde{s}_{11} - s_{12}^2)} \quad \text{(S2.12a)}$$

$$X_{21} = -\frac{F_{12}R_{12}}{s_{11}} - \frac{(s_{11}\widetilde{F}_{12} - s_{12}F_{12})(s_{11}\widetilde{R}_{11} - s_{12}R_{12})}{s_{11}(s_{11}\widetilde{s}_{11} - s_{12}^2)} \quad \text{(S2.12b)}$$

$$Z_{21} = -\frac{F_{12}Q_{12}}{s_{11}} - \frac{(s_{11}\widetilde{F}_{12} - s_{12}F_{12})(s_{11}\widetilde{Q}_{12} - s_{12}Q_{12})}{s_{11}(s_{11}\widetilde{s}_{11} - s_{12}^2)} \quad \text{(S2.12c)}$$

$$G_{11} = \widetilde{g}_{11}^{\sigma} + \frac{F_{12}^2}{2s_{11}} + \frac{(s_{11}\widetilde{F}_{12} - s_{12}F_{12})^2}{2s_{11}(s_{11}\widetilde{s}_{11} - s_{12}^2)}, \quad G_{66} = \widetilde{g}_{66}^{\sigma} \quad \text{(S2.13)}$$

Corresponding Euler-Lagrange equations are

$$2A_1\widetilde{P}_1 + 4A_{11}\widetilde{P}_1^3 + 2A_{12}\widetilde{P}_1\widetilde{P}_2^2 - 2E_{11}\widetilde{\Phi}_1^2\widetilde{P}_1 - 2E_{21}\widetilde{\Phi}_2^2\widetilde{P}_1 - E_{66}\widetilde{\Phi}_1\widetilde{\Phi}_2\widetilde{P}_1 - $$
$$- G_{11}\frac{\partial^2\widetilde{P}_1}{\partial \widetilde{x}_1^2} - 2\left(X_{11}\widetilde{\Phi}_1\frac{\partial\widetilde{\Phi}_1}{\partial \widetilde{x}_1} + X_{21}\widetilde{\Phi}_2\frac{\partial\widetilde{\Phi}_2}{\partial \widetilde{x}_1} + Z_{21}\widetilde{P}_2\frac{\partial\widetilde{P}_2}{\partial \widetilde{x}_1}\right) = \widetilde{E}_1^d \quad \text{(S2.14a)}$$

$$2A_2\widetilde{P}_2 + 4A_{22}\widetilde{P}_2^3 + 2A_{12}\widetilde{P}_1^2\widetilde{P}_2 - 2E_{22}\widetilde{\Phi}_2^2\widetilde{P}_2 - 2E_{12}\widetilde{\Phi}_1^2\widetilde{P}_2 - E_{66}\widetilde{\Phi}_1\widetilde{\Phi}_2\widetilde{P}_1 - G_{66}\frac{\partial^2\widetilde{P}_2}{\partial \widetilde{x}_1^2} + 2Z_{21}\widetilde{P}_2\frac{\partial\widetilde{P}_1}{\partial \widetilde{x}_1} = 0$$
(S2.14b)

$$2B_1\widetilde{\Phi}_1 + 4B_{11}\widetilde{\Phi}_1^3 + 2B_{12}\widetilde{\Phi}_1\widetilde{\Phi}_2^2 - 2E_{11}\widetilde{\Phi}_1\widetilde{P}_1^2 - 2E_{12}\widetilde{\Phi}_1\widetilde{P}_2^2 - E_{66}\widetilde{\Phi}_2\widetilde{P}_1\widetilde{P}_2 - \widetilde{v}_{11}\frac{\partial^2\widetilde{\Phi}_1}{\partial \widetilde{x}_1^2} + 2X_{11}\widetilde{\Phi}_1\frac{\partial\widetilde{P}_1}{\partial \widetilde{x}_1} = 0$$
(S2.14c)

$$2B_2\widetilde{\Phi}_2 + 4B_{22}\widetilde{\Phi}_2^4 + 2B_{12}\widetilde{\Phi}_1^2\widetilde{\Phi}_2 - 2E_{22}\widetilde{\Phi}_2\widetilde{P}_2^2 - 2E_{21}\widetilde{\Phi}_2\widetilde{P}_1^2 - E_{66}\widetilde{\Phi}_2\widetilde{P}_1\widetilde{P}_2 - \widetilde{v}_{66}\frac{\partial^2\widetilde{\Phi}_2}{\partial \widetilde{x}_1^2} + 2X_{21}\widetilde{\Phi}_2\frac{\partial\widetilde{P}_1}{\partial \widetilde{x}_1} = 0$$
(S2.14d)

**Table S2**. List of parameters for polarization and tilt dependent part of the free energy

| Parameter (SI units) | SrTiO$_3$ | |
|---|---|---|
| | **Value** | **Source and notes** |
| $\varepsilon_b$ | 43 | Ref. [3, 4] |
| $\alpha_T^{(P)}$ ($\times 10^6$ m/(F K)) | 0.75 | [5,6] |
| $T_c^{(P)}$ (K) | 30 | [5, 6] |
| $T_q^{(P)}$ (K) | 54 | [5, 6] |
| $a_{11}^{\sigma}$ ($\times 10^9$ m$^5$/(C$^2$F)) | 1.696 | [5, 6] |
| $a_{12}^{\sigma}$ ($\times 10^9$ m$^5$/(C$^2$F)) | 3.918 | [7] |
| $Q_{ij}$ (m$^4$/C$^2$) | $Q_{11}$=0.046, $Q_{12}$= −0.014, $Q_{44}$=0.019 | Recalculated from [5] |
| $\alpha_T^{(\Phi)}$ ($\times 10^{26}$ J/(m$^5$ K)) | 9.1 | [8] |
| $T_c^{(\Phi)}$ (K) | 105 | [8] |
| $T_q^{(\Phi)}$ (K) | 145 | [8] |
| $\beta_{\Phi 11}$ ($\times 10^{50}$ J/m$^7$) | 1.69 | [8] |
| $\beta_{\Phi 12}$ ($\times 10^{50}$ J/m$^7$) | 3.88 | [8] |
| $R_{ij}$ ($\times 10^{18}$ m$^{-2}$) | $R_{11}$=8.7, $R_{12}$= -7.8, $R_{44}$= -18.4 | recalculated from [8] |
| $\eta_{ij}$ ($\times 10^{29}$ (F m)$^{-1}$) | $\eta_{11}^\sigma$=-1.744, $\eta_{12}^\sigma$=-0.755, $\eta_{44}^\sigma$=5.85 | [8] |
| $s_{ij}$ ($\times 10^{-12}$ m$^3$/J) | $s_{11}$=3.52, $s_{12}$= -0.85, $s_{44}$= 7.87 | recalculated from [8, 6] |
| Tilt gradient $v_{ijkl}$ ($10^{10} \times$ J/m$^3$) | $v_{11}$=0.28, $v_{12}$= -7.34, $v_{44}$=7.11 | From [6] |
| $g_{ijkl}$ ($10^{-11} \times$ V·m$^3$/C) | $g_{11}=g_{44}=1$, $g_{12}=0.5$ | Estimation |
| Flexoelectric tensor $F_{ijkl}$ ($10^{-12} \times$ m$^3$/C) | $F_{11}$= -13.80, $F_{12}$= 6.66, $F_{44}$= 8.48 | calculated from the tensor $f_{ij}$ measured by Zubko et al. [9] |

# Appendix S3

## Equations in Φ-P-u representation

For the considered case of 1D distribution near [110] TB the free energy density in the rotated frame is

$$\begin{aligned}
F &= a_1(T)(\widetilde{P}_1^2 + \widetilde{P}_2^2) + \widetilde{a}_{11}^u(\widetilde{P}_1^4 + \widetilde{P}_2^4) + \widetilde{a}_{12}^u(\widetilde{P}_1^2 \widetilde{P}_2^2) - \widetilde{P}_i \widetilde{E}_i^{ext} - \frac{1}{2}\widetilde{P}_1 \widetilde{E}_1^d \\
&- \widetilde{q}_{11}(\widetilde{u}_1 \widetilde{P}_1^2 + \widetilde{u}_2 \widetilde{P}_2^2) - \widetilde{q}_{12}(\widetilde{u}_1 \widetilde{P}_2^2 + \widetilde{u}_2 \widetilde{P}_1^2) - q_{12}\widetilde{u}_3(\widetilde{P}_1^2 + \widetilde{P}_2^2) - \widetilde{q}_{66}\widetilde{u}_6 \widetilde{P}_1 \widetilde{P}_2 \\
&+ \frac{1}{2}\widetilde{c}_{11}(\widetilde{u}_1^2 + \widetilde{u}_2^2) + \frac{1}{2}c_{11}\widetilde{u}_3^2 + \widetilde{c}_{12}\widetilde{u}_1 \widetilde{u}_2 + c_{12}(\widetilde{u}_2 \widetilde{u}_3 + \widetilde{u}_3 \widetilde{u}_1) + \frac{1}{2}\widetilde{c}_{66}\widetilde{u}_6^2 + \frac{1}{2}c_{44}(\widetilde{u}_4^2 + \widetilde{u}_5^2) \\
&+ \beta_1(T)(\widetilde{\Phi}_1^2 + \widetilde{\Phi}_2^2) + \widetilde{\beta}_{11}^u(\widetilde{\Phi}_1^4 + \widetilde{\Phi}_2^4) + \widetilde{\beta}_{12}^u \widetilde{\Phi}_1^2 \widetilde{\Phi}_2^2 \\
&- \widetilde{r}_{11}(\widetilde{u}_1 \widetilde{\Phi}_1^2 + \widetilde{u}_2 \widetilde{\Phi}_2^2) - \widetilde{r}_{12}(\widetilde{u}_1 \widetilde{\Phi}_2^2 + \widetilde{u}_2 \widetilde{\Phi}_1^2) - r_{12}\widetilde{u}_3(\widetilde{\Phi}_1^2 + \widetilde{\Phi}_2^2) - \widetilde{r}_{66}(\widetilde{u}_6 \widetilde{\Phi}_1 \widetilde{\Phi}_2) - \\
&- \widetilde{\eta}_{11}^u(\widetilde{\Phi}_1^2 \widetilde{P}_1^2 + \widetilde{\Phi}_2^2 \widetilde{P}_2^2) - \widetilde{\eta}_{12}^u(\widetilde{\Phi}_1^2 \widetilde{P}_2^2 + \widetilde{\Phi}_2^2 \widetilde{P}_1^2) - \widetilde{\eta}_{66}^u(\widetilde{\Phi}_1 \widetilde{\Phi}_2 \widetilde{P}_1 \widetilde{P}_2) \\
&+ \frac{\widetilde{v}_{11}}{2}\left(\frac{\partial \widetilde{\Phi}_1}{\partial \widetilde{x}_1}\right)^2 + \frac{\widetilde{v}_{66}}{2}\left(\frac{\partial \widetilde{\Phi}_2}{\partial \widetilde{x}_1}\right)^2 + \frac{\widetilde{g}_{11}^u}{2}\left(\frac{\partial \widetilde{P}_1}{\partial \widetilde{x}_1}\right)^2 + \frac{\widetilde{g}_{66}^u}{2}\left(\frac{\partial \widetilde{P}_2}{\partial \widetilde{x}_1}\right)^2 \\
&+ (\widetilde{f}_{11}\widetilde{u}_1 + \widetilde{f}_{12}\widetilde{u}_2 + \widetilde{f}_{12}\widetilde{u}_3)\frac{\partial \widetilde{P}_1}{\partial \widetilde{x}_1} + \widetilde{f}_{66}\widetilde{u}_6 \frac{\partial \widetilde{P}_2}{\partial \widetilde{x}_1}
\end{aligned} \quad (S3.1)$$

Here $\widetilde{E}_i^{ext}$ and $\widetilde{E}_i^d$ are the components of external and depolarization field respectively. In these equations we used the tensor components in the new reference frame for elastic compliances $\widetilde{c}_{11} = (c_{11} + c_{12} + 2c_{44})/2$, $\widetilde{c}_{12} = (c_{11} + c_{12} - 2c_{44})/2$, $\widetilde{c}_{66} = (c_{11} - c_{12})/2$; rotostriction $\widetilde{r}_{11} = (r_{11} + r_{12} + r_{44})/2$, $\widetilde{r}_{12} = (r_{11} + r_{12} - r_{44})/2$, $\widetilde{r}_{66} = (r_{11} - r_{12})$; electrostriction $\widetilde{q}_{11} = (q_{11} + q_{12} + q_{44})/2$, $\widetilde{q}_{12} = (q_{11} + q_{12} - q_{44})/2$, $\widetilde{q}_{66} = (q_{11} - q_{12})$; flexoelectric coefficients $\widetilde{f}_{11} = (f_{11} + f_{12} + 2f_{44})/2$, $\widetilde{f}_{12} = (f_{11} + f_{12} - 2f_{44})/2$, $\widetilde{f}_{66} = (f_{11} - f_{12})/2$.

Using LGD approach we analyze the behavior on the polar ($P_i$) and structural ($\Phi_i$) order parameter components in the presence of ferroelastic surface. Equations of state are:

$$2 b_i \Phi_i + 4 b_{ij}^u \Phi_j^2 \Phi_i - v_{ijkl} \frac{\partial^2 \Phi_k}{\partial x_j \partial x_l} - 2 r_{mjki} u_{mj} \Phi_k - 2 \eta_{klij}^u P_k P_l \Phi_j = 0, \qquad \text{(S3.2a)}$$

$$2 a_i P_i + 4 a_{ijkl}^u P_j P_k P_l - g_{ijkl} \frac{\partial^2 P_k}{\partial x_j \partial x_l} - 2 q_{mjki} u_{mj} P_k - f_{mnil} \frac{\partial u_{mn}}{\partial x_l} - 2 \eta_{ijkl}^u P_j \Phi_k \Phi_l = E_i^d, \qquad \text{(S3.2b)}$$

$$c_{ijkl} u_{kl} - r_{ijkl} \Phi_k \Phi_l + f_{ijkl} (\partial P_k / \partial x_l) - q_{ijkl} P_k P_l = \sigma_{ij}. \qquad \text{(S3.2c)}$$

For considered geometry the elastic solution for strain tensor in Voigt notations (11=1, 22=2, 33=3, 23=4, 13=5, 12=6) has the form:

$$\tilde{u}_{33} = R_{12} \Phi_B^2, \quad \tilde{u}_{22} = \tilde{R}_{11} \frac{\Phi_S^2}{2} + \tilde{R}_{12} \frac{\Phi_S^2}{2}, \qquad \text{(S3.3a)}$$

$$\tilde{u}_{11} = \tilde{R}_{11} \frac{\Phi_S^2}{2} + \tilde{R}_{12} \frac{\Phi_S^2}{2} + \frac{1}{\tilde{c}_{11}} \left( \tilde{q}_{11} \tilde{P}_1^2 + \tilde{q}_{12} \tilde{P}_2^2 + \tilde{r}_{11} \left( \tilde{\Phi}_1^2 - \frac{\Phi_S^2}{2} \right) + \tilde{r}_{12} \left( \tilde{\Phi}_2^2 - \frac{\Phi_S^2}{2} \right) \right) - \frac{\tilde{f}_{11}}{\tilde{c}_{11}} \frac{\partial \tilde{P}_1}{\partial \tilde{x}_1}, \qquad \text{(S3.3b)}$$

$$\tilde{u}_{12} = \frac{1}{2} \left( -\frac{\tilde{f}_{66}}{\tilde{c}_{66}} \frac{\partial \tilde{P}_2}{\partial \tilde{x}_1} + \frac{\tilde{q}_{66}}{\tilde{c}_{66}} \tilde{P}_1 \tilde{P}_2 + \frac{\tilde{r}_{66}}{\tilde{c}_{66}} \tilde{\Phi}_1 \tilde{\Phi}_2 \right), \quad \tilde{u}_{23} = 0, \quad \tilde{u}_{13} = 0. \qquad \text{(S3.3c)}$$

After the substitution of the elastic solution one could get the following system of coupled equations

$$\begin{aligned} & 2 A_1' \tilde{P}_1 + 4 A_{11}' \tilde{P}_1^3 + 2 A_{12}' \tilde{P}_1 \tilde{P}_2^2 - 2 E_{11}' \tilde{P}_1 \tilde{\Phi}_1^2 - 2 E_{21}' \tilde{P}_1 \tilde{\Phi}_2^2 - E_{66}' \tilde{P}_2 \tilde{\Phi}_1 \tilde{\Phi}_2 \\ & - G_{11}' \frac{\partial^2 \tilde{P}_1}{\partial \tilde{x}_1^2} - 2 \left( X_{11}' \tilde{\Phi}_1 \frac{\partial \tilde{\Phi}_1}{\partial \tilde{x}_1} + X_{21}' \tilde{\Phi}_2 \frac{\partial \tilde{\Phi}_2}{\partial \tilde{x}_1} + Z_{21}' \tilde{P}_2 \frac{\partial \tilde{P}_2}{\partial \tilde{x}_1} \right) = \tilde{E}_1^d \end{aligned}, \qquad \text{(S3.4a)}$$

$$\begin{aligned} & 2 A_2' \tilde{P}_2 + 4 A_{22}' \tilde{P}_2^3 + 2 A_{12}' \tilde{P}_2 \tilde{P}_1^2 - 2 E_{12}' \tilde{P}_2 \tilde{\Phi}_1^2 - 2 E_{22}' \tilde{P}_2 \tilde{\Phi}_2^2 - E_{66}' \tilde{P}_1 \tilde{\Phi}_1 \tilde{\Phi}_2 - \\ & - G_{66}' \frac{\partial^2 \tilde{P}_2}{\partial \tilde{x}_1^2} + 2 Z_{21}' \tilde{P}_2 \frac{\partial \tilde{P}_1}{\partial \tilde{x}_1} - X_{66}' \frac{\partial (\tilde{\Phi}_1 \tilde{\Phi}_2)}{\partial \tilde{x}_1} = 0 \end{aligned}, \qquad \text{(S3.4b)}$$

$$\begin{aligned} & 2 B_1' \tilde{\Phi}_1 + 4 B_{11}' \tilde{\Phi}_1^3 + 2 B_{12}' \tilde{\Phi}_1 \tilde{\Phi}_2^2 - 2 E_{11}' \tilde{\Phi}_1 \tilde{P}_1^2 - 2 E_{12}' \tilde{\Phi}_1 \tilde{P}_2^2 - E_{66}' \tilde{P}_1 \tilde{P}_2 \tilde{\Phi}_2 - \\ & - \tilde{v}_{11} \frac{\partial^2 \tilde{\Phi}_1}{\partial \tilde{x}_1^2} + 2 X_{11}' \frac{\partial \tilde{P}_1}{\partial \tilde{x}_1} \tilde{\Phi}_1 + X_{66}' \tilde{\Phi}_2 \frac{\partial \tilde{P}_2}{\partial \tilde{x}_1} = 0 \end{aligned}, \qquad \text{(S3.4c)}$$

$$\begin{aligned} & 2 B_2' \tilde{\Phi}_2 + 4 B_{22}' \tilde{\Phi}_2^3 + 2 B_{12}' \tilde{\Phi}_2 \tilde{\Phi}_1^2 - 2 E_{21}' \tilde{\Phi}_2 \tilde{P}_1^2 - 2 E_{22}' \tilde{\Phi}_2 \tilde{P}_2^2 - E_{66}' \tilde{P}_1 \tilde{P}_2 \tilde{\Phi}_1 - \\ & - \tilde{v}_{66} \frac{\partial^2 \tilde{\Phi}_2}{\partial \tilde{x}_1^2} + 2 X_{21}' \frac{\partial \tilde{P}_1}{\partial \tilde{x}_1} \tilde{\Phi}_2 + X_{66}' \tilde{\Phi}_1 \frac{\partial \tilde{P}_2}{\partial \tilde{x}_1} = 0 \end{aligned}, \qquad \text{(S3.4d)}$$

Here we introduced the designations for renormalized coefficients

$$A_1' = a_1 - \frac{\Phi_S^2}{2} \left( \frac{(q_{11} - q_{12})(r_{11} - r_{12})}{3(c_{11} - c_{12})} + \frac{2(q_{11} + 2 q_{12})(r_{11} + 2 r_{12})}{3(c_{11} + 2 c_{12})} - \frac{(\tilde{r}_{11} + \tilde{r}_{12}) \tilde{q}_{11}}{\tilde{c}_{11}} \right), \qquad \text{(S3.5a)}$$

$$A_2' = a_1 - \frac{\Phi_S^2}{2} \left( \frac{(q_{11} - q_{12})(r_{11} - r_{12})}{3(c_{11} - c_{12})} + \frac{2(q_{11} + 2 q_{12})(r_{11} + 2 r_{12})}{3(c_{11} + 2 c_{12})} - \frac{(\tilde{r}_{11} + \tilde{r}_{12}) \tilde{q}_{12}}{\tilde{c}_{11}} \right) \qquad \text{(S3.5b)}$$

$$B_1' = b_1 - \frac{\Phi_S^2}{2} \left( \frac{(r_{11} - r_{12})^2}{3(c_{11} - c_{12})} + \frac{2(r_{11} + 2 r_{12})^2}{3(c_{11} + 2 c_{12})} + \frac{(\tilde{r}_{11} + \tilde{r}_{12}) \tilde{r}_{11}}{\tilde{c}_{11}} \right), \qquad \text{(S3.5c)}$$

$$B'_2 = b_1 - \frac{\Phi_S^2}{2}\left(\frac{(r_{11}-r_{12})^2}{3(c_{11}-c_{12})} + \frac{2(r_{11}+2r_{12})^2}{3(c_{11}+2c_{12})} + \frac{(\tilde{r}_{11}+\tilde{r}_{12})\tilde{r}_{12}}{\tilde{c}_{11}}\right). \tag{S3.5d}$$

$$A'_{11} = \tilde{a}^u_{11} - \frac{\tilde{q}^2_{11}}{2\tilde{c}_{11}}, \quad A'_{12} = \tilde{a}^u_{12} - \frac{\tilde{q}_{11}\tilde{q}_{12}}{\tilde{c}_{11}} - \frac{\tilde{q}^2_{66}}{2\tilde{c}_{66}}, \quad A'_{22} = \tilde{a}^u_{11} - \frac{\tilde{q}^2_{12}}{2\tilde{c}_{11}}; \tag{S3.6a}$$

$$B'_{11} = \tilde{b}^u_{11} - \frac{\tilde{r}^2_{11}}{2\tilde{c}_{11}} \quad B'_{12} = \tilde{b}^u_{12} - \frac{\tilde{r}_{11}\tilde{r}_{12}}{\tilde{c}_{11}} - \frac{\tilde{r}^2_{66}}{2\tilde{c}_{66}}, \quad B'_{22} = \tilde{b}^u_{11} - \frac{\tilde{r}^2_{12}}{2\tilde{c}_{11}}; \tag{S3.6b}$$

$$E'_{11} = \tilde{\eta}^u_{11} + \frac{\tilde{q}_{11}\tilde{r}_{11}}{\tilde{c}_{11}}, \quad E'_{12} = \tilde{\eta}^u_{12} + \frac{\tilde{q}_{12}\tilde{r}_{11}}{\tilde{c}_{11}}, \quad E'_{21} = \tilde{\eta}^u_{12} + \frac{\tilde{q}_{11}\tilde{r}_{12}}{\tilde{c}_{11}} \quad E'_{22} = \tilde{\eta}^u_{11} + \frac{\tilde{q}_{12}\tilde{r}_{12}}{\tilde{c}_{11}}, \quad E'_{66} = \tilde{\eta}^u_{66} + \frac{\tilde{q}_{66}\tilde{r}_{66}}{\tilde{c}_{66}},$$
$$\tag{S3.7}$$

$$X'_{11} = \frac{\tilde{f}_{11}}{\tilde{c}_{11}}\tilde{r}_{11}, \quad X'_{21} = \frac{\tilde{f}_{11}}{\tilde{c}_{11}}\tilde{r}_{12}, \quad X'_{66} = \frac{\tilde{f}_{66}\tilde{r}_{66}}{\tilde{c}_{66}}; \tag{S3.8}$$

$$Z'_{21} = \frac{\tilde{f}_{11}\tilde{q}_{12}}{\tilde{c}_{11}} - \frac{\tilde{f}_{66}\tilde{q}_{66}}{2\tilde{c}_{66}} \tag{S3.9}$$

$$G'_{11} = \tilde{g}^u_{11} - \frac{\tilde{f}^2_{11}}{\tilde{c}_{11}}, \quad G'_{66} = \tilde{g}^u_{66} - \frac{\tilde{f}^2_{66}}{\tilde{c}_{66}} \tag{S3.10}$$

Despite the formal difference between the renormalized coefficients in Eqs.(S2.14) and (S3.4), actually most of the renormalized coefficients in Φ-P-u and Φ-P-σ representations coincide (like $A_{ij}$ and $A'_{ij}$, $B_{ij}$ and $B'_{ij}$, $E_{ij}$ and $E'_{ij}$, $G_{ij}$ and $G'_{ij}$) as it should be expected. For instance, using Eqs.(S1.3)-(S1.9) after simple but cumbersome algebraic transformations one could show that

$$A_{11} - A'_{11} = \tilde{a}^\sigma_{11} + \frac{Q^2_{12}}{2s_{11}} + \frac{(s_{11}\tilde{Q}_{12} - s_{12}Q_{12})^2}{2s_{11}(s_{11}\tilde{s}_{11} - s^2_{12})} - \tilde{a}^u_{11} + \frac{\tilde{q}^2_{11}}{2\tilde{c}_{11}} =$$
$$= \frac{1}{4}\left(-\frac{(Q_{11}-Q_{12})^2}{3(s_{11}-s_{12})} - 2\frac{(Q_{11}+2Q_{12})^2}{3(s_{11}+2s_{12})} - \frac{Q^2_{44}}{2s_{44}}\right) + \frac{s_{11}\tilde{Q}^2_{12} + \tilde{s}_{11}Q^2_{12} - 2s_{12}\tilde{Q}_{12}Q_{12}}{2(s_{11}\tilde{s}_{11} - s^2_{12})} + \frac{\tilde{q}^2_{11}}{2\tilde{c}_{11}} \equiv 0 \tag{S3.11a}$$

$$A_{22} - A'_{22} = \tilde{a}^\sigma_{11} + \frac{Q^2_{12}}{2s_{11}} + \frac{(s_{11}\tilde{Q}_{11} - s_{12}Q_{12})^2}{2s_{11}(s_{11}\tilde{s}_{11} - s^2_{12})} - \tilde{a}^u_{11} + \frac{\tilde{q}^2_{12}}{2\tilde{c}_{11}} =$$
$$= \frac{1}{4}\left(-\frac{(Q_{11}-Q_{12})^2}{3(s_{11}-s_{12})} - 2\frac{(Q_{11}+2Q_{12})^2}{3(s_{11}+2s_{12})} - \frac{Q^2_{44}}{2s_{44}}\right) + \frac{Q^2_{12}}{2s_{11}} + \frac{(s_{11}\tilde{Q}_{11} - s_{12}Q_{12})^2}{2s_{11}(s_{11}\tilde{s}_{11} - s^2_{12})} + \frac{\tilde{q}^2_{12}}{2\tilde{c}_{11}} \equiv 0 \tag{S3.11b}$$

$$A_{12} - A'_{12} = \tilde{a}^\sigma_{12} + \frac{Q^2_{12}}{s_{11}} + \frac{(s_{11}\tilde{Q}_{12} - s_{12}Q_{12})(s_{11}\tilde{Q}_{11} - s_{12}Q_{12})}{s_{11}(s_{11}\tilde{s}_{11} - s^2_{12})} - \tilde{a}^u_{12} + \frac{\tilde{q}_{11}\tilde{q}_{12}}{\tilde{c}_{11}} + \frac{\tilde{q}^2_{66}}{2\tilde{c}_{66}} =$$

$$\left| \left(3a^\sigma_{11} - \frac{a^\sigma_{12}}{2}\right) - \left(3a^u_{11} - \frac{a^u_{12}}{2}\right) = -\frac{7(Q_{11} - Q_{12})^2}{6(s_{11} - s_{12})} - \frac{(Q_{11} + 2Q_{12})^2}{3(s_{11} + 2s_{12})} + \frac{Q^2_{44}}{4s_{44}} \right|$$

(S3.11c)

$$= -\frac{7(Q_{11} - Q_{12})^2}{6(s_{11} - s_{12})} - \frac{(Q_{11} + 2Q_{12})^2}{3(s_{11} + 2s_{12})} + \frac{Q^2_{44}}{4s_{44}} +$$

$$+\frac{Q^2_{12}}{s_{11}} + \frac{(s_{11}\tilde{Q}_{12} - s_{12}Q_{12})(s_{11}\tilde{Q}_{11} - s_{12}Q_{12})}{s_{11}(s_{11}\tilde{s}_{11} - s^2_{12})} + \frac{\tilde{q}_{11}\tilde{q}_{12}}{\tilde{c}_{11}} + \frac{\tilde{q}^2_{66}}{2\tilde{c}_{66}} \equiv 0$$

However the flexo-like coupling coefficients ($X_{ij}$ and $X'_{ij}$, $Z_{21}$ and $Z'_{21}$) appeared different even after transformations, which is especially obvious in the case of coefficient $X'_{66} \neq 0$ (see Eq.(S3.8)), since one could see from Appendix S2, Eqs.(S2.12a), (S2.12b) that $X_{66} \equiv 0$! The source of the difference lies in the different suppositions used in Appendices S2 and S3 (see the main text). Note that the terms like $\tilde{P}_2 \frac{\partial(\tilde{\Phi}_1 \tilde{\Phi}_2)}{\partial \tilde{x}_1}$ have a profound effect on the structure of twins in AFD systems.

**Appendix S4. Analytical description of the phase diagram with modulated phase induced by the coupling**

Considering only one-component for polarization and tilt vectors, corresponding free energy density (1) and (3) acquires the form:

$$F_b[P,\Phi] = \begin{pmatrix} \beta_1\Phi^2 + \beta_{11}\Phi^4 + \frac{v}{2}\left(\frac{\partial\Phi}{\partial x}\right)^2 + \alpha_1 P^2 + \alpha_{11}P^4 + \frac{g}{2}\left(\frac{\partial P}{\partial x}\right)^2 \\ -\eta P^2\Phi^2 + \frac{\xi}{2}\left(\frac{\partial P}{\partial x}\Phi^2 - P\frac{\partial(\Phi^2)}{\partial x}\right) \end{pmatrix}$$

(S4.1)

The coupling coefficient $\xi = \xi^\sigma + F \cdot r$ is already renormalized by the roto-flexoelectric effect. The free energy (S4.1) is stable at high values of order parameters only under the conditions: $\alpha_{11} > 0$, $\beta_{11} > 0$, $2\sqrt{\alpha_{11}\beta_{11}} - \eta > 0$. Note, that depolarization field does not affect directly the polarization components $P_i(x_{j \neq i})$, since $\frac{\partial P_i(x_{j \neq i})}{\partial x_i} \equiv 0$. In particular $\frac{\partial P_3(x_1)}{\partial x_3} = 0$, and the case is considered below.

**Stable homogeneous phases.** In the homogeneous case the free energy functional (S4.1) minimization with respect to the order parameters gives

$$2\beta_1\Phi + 4\beta_{11}\Phi^3 - 2\eta\Phi P^2 = 0,$$  (S4.2a)

$$2\alpha_1 P + 4\alpha_{11} P^3 - 2\eta \Phi^2 P = 0. \qquad (S4.2b)$$

The possible phases with homogeneous distribution of the order parameters are listed below

**1. Para phase** with $P = \Phi = 0$ and free energy $F = 0$. This phase is stable at $\alpha_1 > 0$ and $\beta_1 > 0$

**2. AFD-phase** with $\Phi = \pm\sqrt{\dfrac{-\beta_1}{2\beta_{11}}}$, $P = 0$ and Free energy $F = \dfrac{-\beta_1^2}{4\beta_{11}}$. This phase exists and is stable under the conditions $\beta_1 < 0$ and $2\alpha_1 + \eta(\beta_1/\beta_{11}) > 0$.

**3. FE-AFD phase** with $P = \sqrt{-\dfrac{2\alpha_1 + \eta(\beta_1/\beta_{11})}{4\alpha_{11} - (\eta^2/\beta_{11})}}$, $\Phi = \sqrt{-\dfrac{2\beta_1 + \eta(\alpha_1/\alpha_{11})}{4\beta_{11} - (\eta^2/\alpha_{11})}}$. Free energy is

$$F = -\frac{\beta_{11}\alpha_1^2 + \alpha_{11}\beta_1^2 + \eta\alpha_1\beta_1}{4\alpha_{11}\beta_{11} - \eta^2}.$$ This phase exists and is stable under the conditions

$2\alpha_1 + \eta(\beta_1/\beta_{11}) < 0$, $2\beta_1 + \eta(\alpha_1/\alpha_{11}) < 0$, $2\sqrt{\alpha_{11}\beta_{11}} > -\eta$.

**4. FE phase** with $\Phi = 0$, $P = \sqrt{\dfrac{-\alpha_1}{2\alpha_{11}}}$, Free energy is $F = \dfrac{-\alpha_1^2}{4\alpha_{11}}$. This phase exists and is stable under the conditions $\alpha_1 < 0$, $2\beta_1 + \eta(\alpha_1/\alpha_{11}) > 0$.

Initial free energy

**Stable modulated phase.** Modulated phase could exist near the region between the curves $2\alpha_1 + \eta(\beta_1/\beta_{11}) = 0$ and $2\beta_1 + \eta(\alpha_1/\alpha_{11}) = 0$. In the inhomogeneous case the free energy functional (S4.1) minimization with respect to the order parameters gives

$$2\beta_1 \Phi + 4\beta_{11} \Phi^3 - \nu \frac{\partial^2 \Phi}{\partial x^2} + 2\xi \Phi \frac{\partial P}{\partial x} - 2\eta \Phi P^2 = 0, \qquad (S4.3a)$$

$$2\alpha_1 P + 4\alpha_{11} P^3 - g \frac{\partial^2 P}{\partial x^2} - 2\eta \Phi^2 P - 2\xi \Phi \frac{\partial \Phi}{\partial x} = 0. \qquad (S4.3b)$$

Let us look for the order parameters distribution in the modulated phase onset in harmonic approximation

$$P = \delta P \sin(kx), \qquad \Phi = \Phi_0 - \delta\Phi \cos(kx) \qquad (S4.4)$$

Wave number $k$, "base" $\Phi_0$, amplitudes $\delta P$ and $\delta\Phi$ are variational parameters. After the substitution of Eq.(S4.4) in the potential (S4.1) and integration over a period corresponding smooth part is

$$\widetilde{F}_b[k,\delta P,\delta\Phi,\Phi_0] = \begin{pmatrix} \left(\dfrac{\beta_1}{2} + \dfrac{\nu}{4}k^2\right)\delta\Phi^2 + \dfrac{3\beta_{11}}{8}\delta\Phi^4 + 3\beta_{11}\delta\Phi^2\Phi_0^2 + \beta_1\Phi_0^2 + \beta_{11}\Phi_0^4 + \\ \left(\dfrac{\alpha_1}{2} + \dfrac{g}{4}k^2\right)\delta P^2 + \dfrac{3\alpha_{11}}{8}\delta P^4 - \dfrac{\eta}{8}\left(\delta\Phi^2 + 4\Phi_0^2\right)\delta P^2 - \xi k\Phi_0\delta\Phi\delta P \end{pmatrix} \qquad (S4.5)$$

Minimization with respect to $k$, $\Phi_0$, $\delta P$ and $\delta\Phi$ leads to the coupled system of algebraic equations:

$$\left(\frac{g}{2}\delta P^2 + \frac{v}{2}\delta\Phi^2\right)k - \xi\Phi_0\delta\Phi\delta P = 0, \qquad (S4.6a)$$

$$\left(\alpha_1 + \frac{g}{2}k^2 - \frac{\eta}{4}(\delta\Phi^2 + 4\Phi_0^2)\right)\delta P + \frac{3\alpha_{11}}{2}\delta P^3 - \xi k\Phi_0\delta\Phi = 0, \qquad (S4.6b)$$

$$\left(\beta_1 + \frac{v}{2}k^2 + 6\beta_{11}\Phi_0^2 - \frac{\eta}{4}\delta P^2\right)\delta\Phi + \frac{3\beta_{11}}{2}\delta\Phi^3 - \xi k\Phi_0\delta P = 0, \qquad (S4.6c)$$

$$\left(2\beta_1 + 6\beta_{11}\delta\Phi^2 - \eta\delta P^2\right)\Phi_0 + 4\beta_{11}\Phi_0^3 - \xi k\delta\Phi\delta P = 0. \qquad (S4.6d)$$

Assuming that $|\Phi_0| \gg |\delta\Phi|$ and polarization amplitude is small $\delta P$ at the modulate phase boundary, we derive the approximate solution of the system (S4.6):

$$k = \frac{2\xi\Phi_0\delta\Phi\delta P}{g\delta P^2 + v\delta\Phi^2}, \quad \delta\Phi \approx \frac{\xi k\Phi_0\delta P}{\beta_1 + \frac{v}{2}k^2 + 6\beta_{11}\Phi_0^2}, \quad \delta P \approx \frac{\xi k\Phi_0\delta\Phi}{\alpha_1 + \frac{g}{2}k^2 - \eta\Phi_0^2}, \quad \Phi_0 \approx \pm\sqrt{\frac{-\beta_1}{2\beta_{11}}} \quad (S4.7)$$

So the wave number $k$ should obey biquadratic equation:

$$\xi^2 k^2 \Phi_0^2 \approx \left(\beta_1 + \frac{v}{2}k^2 + 6\beta_{11}\Phi_0^2\right)\left(\alpha_1 + \frac{g}{2}k^2 - \eta\Phi_0^2\right) \qquad (S4.8a)$$

Using approximation for $\Phi_0$, Eq.(S4.8a) can be rewritten as:

$$\left(\frac{v}{2}k^2 - 2\beta_1\right)\left(\alpha_1 + \frac{g}{2}k^2 + \frac{\eta\beta_1}{2\beta_{11}}\right) + \xi^2\frac{\beta_1}{2\beta_{11}}k^2 = 0 \iff ak^4 + bk^2 + c = 0 \qquad (S4.8b)$$

$$a = \frac{gv}{4}, \quad b = \frac{-4g\beta_{11}\beta_1 + v(2\beta_{11}\alpha_1 + \eta\beta_1) + 2\xi^2\beta_1}{4\beta_{11}}, \quad c = -\beta_1\frac{(2\alpha_1\beta_{11} + \eta\beta_1)}{2\beta_{11}} \qquad (S4.8c)$$

The solution is:

$$k^2 = \frac{-b \pm \sqrt{b^2 - 4ac}}{2a} \qquad (S4.8d)$$

For considered case $a > 0$, $c > 0$, since modulation appears in the AFD phase that is stable under the conditions $\beta_1 < 0$ and $2\alpha_1 + \eta(\beta_1/\beta_{11}) > 0$. So, the solution (S4.8d) exists under the conditions $b < 0$ and $b^2 \geq 4ac$, namely $b \leq -2\sqrt{ac}$:

$$\xi^2 > 2g\beta_{11} - v(\beta_{11}(\alpha_1/\beta_1) + \eta/2) + \sqrt{-4gv\beta_{11}(\beta_{11}(\alpha_1/\beta_1) + \eta/2)}. \qquad (S4.9)$$

In fact Eq.(S4.9) determines the critical value of the coupling $\xi$ strength. Modulated phase boundary can be determined from the condition $b = -2\sqrt{ac}$, that immediately leads to the equation for the phase boundary and wave vector at the boundary:

$$\xi^2 = 2g\beta_{11} - v(\beta_{11}(\alpha_1/\beta_1) + \eta/2) + \sqrt{-4g\beta_{11}v(\beta_{11}(\alpha_1/\beta_1) + \eta/2)}$$

$$\equiv \beta_{11}\left(\sqrt{2g} + \sqrt{-\frac{v}{2\beta_1}\left(2\alpha_1 + \eta\frac{\beta_1}{\beta_{11}}\right)}\right)^2 \quad \text{(S4.10a)}$$

$$k = \sqrt{\frac{-b}{2a}} = \sqrt{-\frac{v(2\beta_{11}\alpha_1 + \eta\beta_1) + 2\xi^2\beta_1 - 4g\beta_{11}\beta_1}{2\beta_{11}gv}} \quad \text{(S4.10b)}$$

Dimensionless variables and parameters used in numerical simulations are: $\frac{\Phi}{\Phi_S}$, $\frac{P}{P_S}$, $\tilde{x} = x/L_\Phi$,

where $\Phi_S = \sqrt{\beta_T T_\Phi(0)/\beta_{11}}$, $P_S = \Phi_S\sqrt{\beta_{11}/\alpha_{11}}$, $L_\Phi = \sqrt{\frac{v}{\beta_T T_\Phi(0)}}$ and $\eta^* = \frac{\eta}{\sqrt{\alpha_{11}\beta_{11}}}$,

$\xi^* = \frac{\xi}{\sqrt{v\sqrt{\alpha_{11}\beta_{11}}}}$, $\Delta = \frac{\alpha_T\sqrt{\beta_{11}}}{\beta_T\sqrt{\alpha_{11}}}$, $g^* = \frac{g\sqrt{\beta_{11}}}{v\sqrt{\alpha_{11}}}$.

Figure S1 illustrates the phases evolution in dependence on the coupling constants $\xi^*$ and $\eta^*$ calculated for different renormalized temperatures $t = -1.5$, $-1$, $-0.6$.

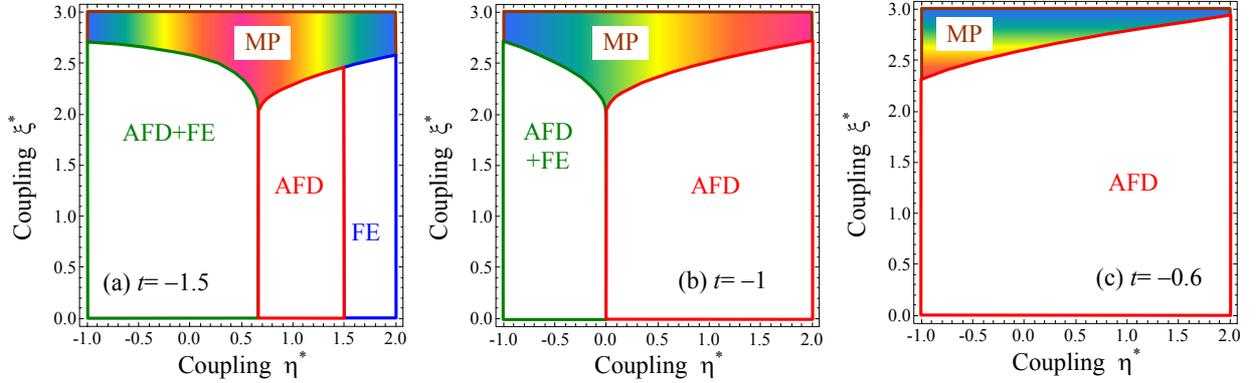

**Figure S1.** Phases evolution in dependence on the coupling constants $\xi^*$ and $\eta^*$ calculated for $\Delta = 0.5$, $g^* = 8$ and different renormalized temperatures $t = -1.5$ **(a)**, $-1$ **(b)**, $-0.6$ **(c)**. Filled regions indicate the MP absolute stability.

### Appendix S5. Estimation of the coupling strength for EuTiO$_3$

Equation (S4.8b) can be identically rewritten as

$$\xi^2 = -\frac{2\beta_{11}}{k^2\beta_1}\left(\frac{v}{2}k^2 - 2\beta_1\right)\left(\alpha_1 + \frac{g}{2}k^2 + \frac{\eta\beta_1}{2\beta_{11}}\right), \quad \text{(S5.1a)}$$

or

$$\xi^2 = \nu g \frac{\left(1 + L_\Phi^2 k^2\right)\left(1 + L_P^2 k^2\right)}{4 L_P^2 L_\Phi^2 k^2 \Phi_0^2}. \tag{S5.1b}$$

Here we introduced the polar and structural correlation lengths as $L_P = \sqrt{g(2\alpha_1 + \eta(\beta_1/\beta_{11}))^{-1}}$ and $L_\Phi = \sqrt{\nu/(-4\beta_1)}$, and bulk expression for antiferrodistortive order parameter $\Phi_0 \approx \pm\sqrt{-\beta_1/2\beta_{11}}$ (tilt).

**Material parameters of EuTiO₃.** Coefficient $\alpha_1(T)$ depends on the temperature $T$ in accordance with Barrett law [10] $\alpha_1(T) = \alpha_T^{(P)}\left(T_q^{(P)}/2\right)\left(\coth\left(T_q^{(P)}/2T\right) - \coth\left(T_q^{(P)}/2T_c^{(P)}\right)\right)$. Coefficient $\alpha_T^{(P)} = 0.98\times10^6$ m/(F K), $T_q^{(P)} \approx 230$ K is the called quantum vibration temperature, $T_c^{(P)} \approx -133.5$ K is the "effective" Curie temperature corresponding to polar soft mode in bulk EuTiO₃ [11, 12]. To account for the experiment and Barrett law the coefficient $\beta_1(T)$ depends on temperature as $\beta_1(T) = \beta_T^{(\Phi)}\left(T_q^{(\Phi)}/2\right)\left(\coth\left(T_q^{(\Phi)}/2T\right) - \coth\left(T_q^{(\Phi)}/2T_c^{(\Phi)}\right)\right)$, where $\beta_T^{(\Phi)} = 1.96\times10^{26}$ J/(m⁵ K), $T_q^{(\Phi)} \approx 205$ K, $T_c^{(\Phi)} \approx 290$ K [13, 14]. Other parameters in the functional (1) can be recalculated [15] as polarization gradient coefficient $g = 10^{-11}$ V·m³/C, tilt gradient nonliear $\nu = \nu_{11} = 0.28\times10^{10}$ J/m³ or $\nu = \nu_{44} = 7.34\times10^{10}$ J/m³ depending on the tilt orientation $\vec{\Phi}$, coefficients $\beta_{11} = 0.436\times10^{50}$ J/m⁷, $\alpha_{11} = 1.6 \times10^9$ m⁵/(C²F), biquadratic coupling coefficient $\eta = \eta_{11} = 2.23\times10^{29}$ (F m)⁻¹ or $\eta = \eta_{12} = -0.85\times10^{29}$ (F m)⁻¹ depending on the $\vec{\Phi}$-orientation.

**Appendix S6. Polarization and tilt at the domain walls between domains with different tilt vectors: emergence of spatially modulated structures.**

Let us consider the structural component of free energy as the LGD polynomial with respect to the structural and polar order parameters as

$$F = \beta_1(T)\Phi^2 + \beta_{11}\Phi^4 + \frac{\nu}{2}\left(\frac{\partial \Phi}{\partial x}\right)^2 + \alpha_1(T)P^2 + \alpha_{11}P^4 + \frac{g}{2}\left(\frac{\partial P}{\partial x}\right)^2 - \eta P^2\Phi^2 + \xi\frac{\partial P}{\partial x}\Phi^2 \tag{S6.1}$$

where $\beta_1(T) = \beta_T T_{q\Phi}\left(\coth(T_{q\Phi}/T) - \coth(T_{q\Phi}/T_\Phi)\right)$ is the temperature dependent coefficient and $T_\Phi$ is the corresponding transition temperature. Here $\alpha_1(T) = \alpha_T T_{qP}\left(\coth(T_{qP}/T) - \coth(T_{qP}/T_P)\right)$, where $T_P$ is the corresponding Curie temperature, and we suppose that $T_\Phi > T_P$ and external electric field to be zero. The homogeneous part of Eq. (S6.1) is rather complex indeed, but was studied in detail earlier (see e.g. paper by Balashova and Tagantsev who studied the system under external electric field [16] and refs. therein).

Equations of state could be obtained from Eqs.(1-4) in the following form

$$2\alpha_1(T)P + 4\alpha_{11}P^3 - 2\eta\Phi^2 P - \xi\frac{\partial(\Phi^2)}{\partial x} - g\frac{\partial^2 P}{\partial x^2} = 0, \quad (S6.2a)$$

$$2\beta_1(T)\Phi + 4\beta_{11}\Phi^3 - 2\eta P^2\Phi + 2\xi\frac{\partial P}{\partial x}\Phi - \nu\frac{\partial^2 \Phi}{\partial x^2} = 0. \quad (S6.2b)$$

Using characteristic values

$$\Phi_b = \sqrt{\frac{\beta_T T_{q\Phi}}{2\beta_{11}}}, \quad P_b = \sqrt{\frac{\beta_T T_{q\Phi}}{2\sqrt{\beta_{11}\alpha_{11}}}}, \quad L_{\Phi 0} = \sqrt{\frac{\nu}{2\beta_T T_{q\Phi}}}, \quad (S6.3)$$

we could introduce dimensionless order variables:

$$\Phi/\Phi_b = f, \quad P/P_b = p, \quad x/L_{\Phi 0} = \tilde{x} \quad (S6.4)$$

and dimensionless parameters:

$$\frac{\eta}{\sqrt{\alpha_{11}\beta_{11}}} \equiv \eta^*, \quad \frac{\alpha_T\sqrt{\beta_{11}}}{\beta_T\sqrt{\alpha_{11}}} \equiv \Delta, \quad \frac{\xi}{2\sqrt{\nu\sqrt{\alpha_{11}\beta_{11}}}} \equiv \xi^*, \quad \frac{g\sqrt{\beta_{11}}}{\nu\sqrt{\alpha_{11}}} \equiv g^*. \quad (S6.5)$$

and could get free energy in the dimensionless form

$$\frac{\beta_{11}F}{(\beta_T T_{q\Phi})^2} = \frac{(\coth(T_{q\Phi}/T) - \coth(T_{q\Phi}/T_\Phi))}{2}f^2 + \frac{1}{4}f^4 + \frac{1}{2}\left(\frac{\partial f}{\partial \tilde{x}}\right)^2 +$$
$$+ \frac{\Delta}{2}(\coth(T_{qP}/T) - \coth(T_{qP}/T_P))p^2 + \frac{1}{4}p^4 + \frac{g^*}{2}\left(\frac{\partial p}{\partial \tilde{x}}\right)^2 - \frac{\eta^*}{4}p^2 f^2 + \xi^* f^2 \frac{\partial p}{\partial \tilde{x}} \quad (S6.6)$$

There are two case depending on the tilt orientation $\vec{\Phi}$ with respect to the direction of modulation vector $\vec{k} \uparrow\uparrow x$. It should be noted that $\xi = \xi_{11}$ for the case $\vec{k} \uparrow\uparrow \vec{\Phi}$ when $\nu = \nu_{11}$ and $\eta = \eta_{11}$, meanwhile $\xi = \xi_{12}$ for the case $\vec{k} \perp \vec{\Phi}$ when $\nu = \nu_{44}$ and $\eta = \eta_{12}$.